\documentclass[journal,twoside]{IEEEtran}
\IEEEoverridecommandlockouts
\usepackage[noadjust]{cite}

\usepackage{amsmath,amssymb,amsfonts}
\usepackage{amsthm}
\usepackage{enumitem}
\usepackage{graphicx}
\usepackage{textcomp}
\usepackage{xcolor}
\usepackage[colorlinks,citecolor=blue,linkcolor=blue,urlcolor=magenta]{hyperref}
\usepackage{booktabs}
\usepackage{cases}
\allowdisplaybreaks[4]
\usepackage{siunitx}
\usepackage{upgreek}
\usepackage{stfloats}
\usepackage{balance}
\usepackage[smallerops,cmintegrals]{newtxmath}
\DeclareMathAlphabet{\mathcal}{OMS}{cmsy}{m}{n}
\usepackage{orcidlink}
\usepackage{upgreek}
\usepackage{physics}
\usepackage{acronym}
\usepackage{algorithm,algorithmic}

\acrodef{ISAC}{integrated sensing and communications}
\acrodef{CRB}{Cram\'er-Rao bound}
\acrodef{SISO}{single-input single-output}
\acrodef{MIMO}{multiple-input multiple-output}
\acrodef{FIM}{Fisher information matrix}
\acrodef{Rx}{receiver}
\acrodef{Tx}{transmitter}
\acrodef{PDF}{probability density function}
\acrodef{PMF}{probability mass function}
\acrodef{MMSE}{minimum mean square error}
\acrodef{LMMSE}{linear \ac{MMSE}}
\acrodef{AoA}{angle of arrival}
\acrodef{BA}{Blahut-Arimoto}
\acrodef{GP}{general power}
\acrodef{MA}{maximum amplitude}
\acrodef{CM}{constant modulus}
\acrodef{KKT}{Karush-Kuhn-Tucker}
\acrodef{SNR}{signal-to-noise-ratio}

\theoremstyle{remark}
\newtheorem{theorem}{Theorem}
\newtheorem{lemma}{Lemma}
\newtheorem{corollary}{Corollary}
\newtheorem{remark}{Remark}
\newtheorem{proposition}{Proposition}

\newcommand{\Y}{{\boldsymbol{\mathsf Y}}}
\newcommand{\x}{{\boldsymbol{\mathsf x}}}
\newcommand{\W}{\boldsymbol{W}}
\newcommand{\s}{\mathsf{s}}

\newcommand{\yr}{\boldsymbol{y}}

\newcommand{\Yc}{\Y_{\mathrm{c}}}
\newcommand{\Ys}{\Y_{\mathrm{s}}}

\newcommand{\Zc}{\boldsymbol{\mathsf Z}_{\mathrm{c}}}
\newcommand{\Zs}{\boldsymbol{\mathsf Z}_{\mathrm{s}}}
\newcommand{\hc}{\boldsymbol{h}_{\mathrm{c}}}
\newcommand{\hs}{\boldsymbol{\mathsf h}_{\mathrm{s}}}
\newcommand{\ys}{\boldsymbol{\mathsf y}_{\mathrm s}}
\newcommand{\yc}{\boldsymbol{\mathsf y}_{\mathrm c}}
\newcommand{\zc}{\boldsymbol{\mathsf z}_{\mathrm c}}
\renewcommand{\u}{\boldsymbol{\mathsf u}}
\renewcommand{\v}{\boldsymbol{\mathsf{v}}}

\newcommand{\noisevar}{\sigma^2_{\mathsf{z}}}

\newcommand{\I}{\boldsymbol{I}}
\newcommand{\Fr}{\boldsymbol{\mathsf F}}
\newcommand{\Fp}{\boldsymbol{F}_{\mathrm{p}}}

\newcommand{\boldeta}{\boldsymbol{\upeta}}

\newcommand{\CRBm}{\overline{\boldsymbol{\mathsf{CRB}}}}

\newcommand{\Lhs}{L_{\mathsf{h}_{\mathrm s}}}
\newcommand{\Lhseff}{\tilde{L}_{\mathsf{h}_{\mathrm s}}}
\newcommand{\Lhc}{L_{h_{\mathrm c}}}
\newcommand{\Ln}{L_{\upeta}}

\newcommand{\boldSigma}{\boldsymbol{\mathit\Sigma}}
\newcommand{\boldLambda}{\boldsymbol{\mathit\Lambda}}

\newcommand{\trans}{\top}
\newcommand{\herm}{\dag}

\newcommand{\vect}{\mathrm{vec}}
\DeclareMathOperator{\diag}{\mathrm{diag}}
\DeclareMathOperator{\supp}{\mathrm{supp}}
\DeclareMathOperator{\card}{\mathrm{card}}
\newcommand{\diff}{\mathrm{d}}
\DeclareMathOperator{\mmse}{{\mathrm{mmse}}}
\DeclareMathOperator{\lmmse}{{\mathrm{lmmse}}}
\DeclareMathOperator{\noz}{{\mathrm{noz}}}

\newcommand{\Px}{\mathbb{P}_{\x}}
\newcommand{\Ps}{\mathbb{P}_{\s}}
\newcommand{\Psopt}{\mathbb{P}_{\s,\alpha}^\star}

\newcommand{\Rate}{\mathcal{R}}
\newcommand{\MI}{\mathcal{I}}
\newcommand{\Fobj}{\mathcal{F}_\alpha}
\newcommand{\SNRc}{\gamma_{\mathrm c}}
\newcommand{\Ku}{\overline{K_{\alpha}}}
\newcommand{\Kl}{\underline{K_{\alpha}}}

\newcommand{\alphas}{\alpha_{\mathrm s}}
\newcommand{\alphac}{\alpha_{\mathrm c}}

\newcommand{\Acm}{A_{\mathrm{CM}}}
\newcommand{\Ama}{A_{\mathrm{MA}}}

\renewcommand{\eqref}[1]{\textcolor{blue}{(\ref{#1})}}

\begin{document}
	
\title{On Pareto-Optimal Estimation-Information Limits-Achieving Distributions for Integrated Sensing and Communications}

\author{
	\IEEEauthorblockN{Zi-Jie Wang$^{\orcidlink{0009-0009-6702-1811}}$} 
	
	\thanks{Zi-Jie Wang is with the Global College, Shanghai Jiao Tong University, Shanghai 200240, China, and also with the Department of Electrical and Computer Engineering, National University of Singapore, Singapore 117583 (e-mail: zijie.wang@u.nus.edu).}
}

\maketitle

\begin{abstract}
	Integrated sensing and communications (ISAC) systems enable simultaneous data transmission and environmental sensing using a common waveform. Characterizing their fundamental estimation-information performance limits is of core interest in information theory. In this paper, the achievable data rate for communications and the Miller-Chang-type Cram\'er-Rao bound (CRB) on sensing distortion are adopted as ISAC performance metrics, both of which are functionals of the ISAC waveform distribution. Our aim is to characterize the optimal input distributions across the entire CRB-Rate Pareto boundary, corresponding to different sensing-communication tradeoff preferences. Three types of power constraint on input waveform are considered, namely, i) general power (GP), ii) maximum amplitude (MA), and iii) constant modulus (CM) constraints. Under these constraints, Pareto-optimal input distributions are shown to exhibit a common structural property: the directional component is uniformly distributed over the complex unit sphere; the power distribution is purely discrete (excluding the communication-optimal point under GP, where it follows a Gamma distribution induced by Gaussian inputs). We further investigate and characterize their Pareto-optimal support structures.
	Under GP, there necessarily exists a non-empty region of the Pareto boundary over which the optimal power distributions possess countably infinitely many mass points. Under MA, the optimal power distributions have finite mass points across the entire Pareto boundary. Sensing-communication preference dependent lower and upper bounds on this finite number are further derived. For CM, it attains a trivial CRB-Rate point instead of a complete Pareto boundary. 
	Finally, numerical algorithms are developed under GP and MA to trace their Pareto boundaries and determine corresponding optimal power distributions based on their respective support characteristics. From both analytical and numerical results, we conclude sensing prefers deterministic power while communication benefits from a richer distribution of power levels, thereby revealing the sensing-communication waveform uncertainty tradeoff in the fundamental power domain.
\end{abstract}

\section{Introduction}

\IEEEPARstart{I}{ntegrated} sensing and communications (ISAC) has emerged as a pivotal paradigm for next-generation wireless networks, enabling simultaneous information transmission and environmental sensing through shared spectral, hardware, and waveform resources \cite{FanLiu6G,LiuAn_survey,signal_processing_technique_ISAC}. Integrating these two functionalities into a unified system, ISAC has the potential to improve spectral efficiency and facilitate a wide range of emerging applications, such as wireless localization and positioning \cite{wang2025device}, environmental monitoring \cite{Zhao2026EnvironmentalISAC}, and intelligent transportation \cite{Cheng2022VehicularISAC}. 

Nevertheless, the fundamentally different requirements of sensing and communications give rise to an inherent performance tradeoff within ISAC systems. To this end, extensive studies have investigated joint sensing-communication performance limits,  typically characterized by estimation-theoretic bounds and information-theoretic metrics, respectively \cite{LiuAn_survey,signal_processing_technique_ISAC}. In \cite{Chiriyath2017RadarCommunicationsConvergence,TVT_BA}, the achievable data rate and the estimation rate, as an analogy to the data rate, are adopted as communication and sensing metrics, respectively. By allocating orthogonal time and frequency resources between sensing and communications, inner performance bounds are derived to characterize the performance of ISAC systems under different setups. Moving beyond orthogonal resource allocation, which treats sensing and communications as competing parties and manages their interference, the pioneering works \cite{CRB_FanLiu_TSP,CRB_hua,CRB_BF} propose to simultaneously optimize sensing and communications functionalities via a unified, common \ac{MIMO} ISAC waveform. Specifically, the celebrated \ac{CRB} for sensing-related parameter estimation is adopted as the sensing metric. The CRB benchmarks a lower bound for estimation distortion using any unbiased or weakly biased estimators \cite[Chapter 3]{Kay1993Fundamentals}. These works have addressed the sensing-communication tradeoff based on covariance reshaping: the sensing CRB is minimized under communication signal-to-interference-plus-noise-ratio \cite{CRB_FanLiu_TSP}, while the achievable rate --- given by the $\log\det(\cdot)$ formula under Gaussian inputs --- is maximized under CRB constraints \cite{CRB_hua, CRB_BF}.

Most of these studies, while insightful, are convex programs over the transmit covariance matrix, and thus characterize the ISAC Pareto optimality only at the level of second-order statistics. More specifically, the sensing performance is evaluated by treating the transmitted waveform, or its sample covariance matrix, as deterministic, thereby overlooking the inherent randomness of information-bearing ISAC signals. On the other hand, the communication performance is accessed with Shannon's $\log\det(\cdot)$ expression, which presumes Gaussian input, while the optimality of Gaussian signaling for ISAC has not been established by then in general. Such treatments do not fully capture the statistical dependence of both sensing and communication performance on the distribution of the transmitted ISAC waveform, potentially limiting the characterization of their fundamental tradeoff and yielding loose bounds on true ISAC performance limits.

More recently, two foundational works \cite{Xiong_TIT,wang_2026_TIT} have explicitly incorporated waveform randomness into the characterization of fundamental estimation-information limits for ISAC systems under Gaussian MIMO channels with average input power constraints. Specifically, \cite{Xiong_TIT} investigated the sensing- and communication-optimal endpoints on the CRB-Rate Pareto front under random input. By investigating the ISAC performance and corresponding achieving strategies at these two points, they successfully reveal the deterministic-random tradeoff associated with waveform distribution and the subspace tradeoff governed by channel characteristics. Building upon this framework, \cite{wang_2026_TIT} further adopted average \ac{MMSE} (e.g., the MMSE averaged according to the statistical distribution of the input waveform) and the achievable rate (e.g., mutual information normalized by transmission/observation time) as sensing and communication metrics, respectively. Under this setup, both metrics are \emph{functionals} of the distribution of the input ISAC waveform. They further characterize the entire \ac{MMSE}-Rate Pareto boundary by formulating and solving functional optimizations. Remarkably, it is established that, except at the communication-optimal endpoint, the Pareto boundary generally cannot be achieved by any continuous input distributions with full support. This result theoretically excludes Gaussian from being ISAC-optimal in general \ac{MIMO} channels, highlighting the non-trivial statistical structure of Pareto-optimal ISAC signaling. 

While \ac{BA}-type numerical algorithms have been developed in \cite{wang_2026_TIT} to approximately determine the optimal input distributions along the Pareto boundary, their computational complexity grows exponentially with the input dimension, rendering them computationally prohibitive for MIMO ISAC systems. In addition, a systematic theoretical characterization of the structural properties of Pareto-optimal input distributions, particularly their support structures and dependence on the sensing-communication tradeoff preference, remains largely unexplored. One major challenge arises from the spatial coupling inherent in MIMO channels, which substantially complicates the theoretical characterization of the joint distribution of high-dimensional ISAC waveforms. 

To isolate the impact of waveform randomness in determining the fundamental ISAC limits from the spatial coupling effects, 	we consider an ISAC system in which a high-dimensional time-domain waveform is transmitted through a single antenna and observed by multiple receive antennas. This paper's organization and theoretical results are summarized as follows.

\begin{itemize}
	\item In Section \ref{sec:system_model}, we present our system model. Specifically, we consider three types of power constraints on the input waveform, namely, \ac{GP}, \ac{MA} \cite{SMITH1971,Dytso2020}, and \ac{CM} constraints. \ac{GP} refers to the average power constraints considered in most studies \cite{wang_2026_TIT,Xiong_TIT,CRB_hua}, while \ac{MA} and \ac{CM} account for practical effects such as hardware limitations on output power. In addition, we choose the Miller-Chang \cite{Miller1978MCRB} type Bayesian \ac{CRB} as the sensing metric \cite{Xiong_TIT} to assess the performance for general parameter estimation tasks, and derive its closed-form, non-matrix expression.
	\item In Section \ref{sec:results}, we establish our main theoretical results. We first show that the Pareto-optimal distributions of the ISAC waveform share a common structure among different power constraints. The directional distribution is uniform over the complex unit sphere, while the power distribution is purely discrete (except at the communication-optimal point under GP). For GP and MA, we derive the \ac{KKT} condition on the power distribution for Pareto optimality; we establish that the sensing-optimal power distribution is a single-mass-point distribution concentrated at the maximum admissible average and peak power, respectively.
	\begin{itemize}
		\item For GP, we show that there exists a non-empty region of the Pareto boundary adjacent to the communication-optimal point, in which the optimal power distribution necessarily possesses countably infinitely many mass points. A sufficient threshold on sensing-communication preference is derived to characterize this region. 
		\item For MA, the Pareto-optimal power distributions are shown to possess finitely many mass points across the entire Pareto boundary. Preference-dependent lower and upper bounds on this finite number are derived. Both bounds are non-decreasing when communication is more favored. 
		\item Under CM, the achievable region has a single Pareto-optimal point.
		\item In addition, we develop separate numerical algorithms to evaluate the Pareto-optimal ISAC performance and power distributions under GP and MA.
	\end{itemize}
	\item In Section \ref{sec:case_study}, we consider \ac{AoA} estimation as the sensing task of interest. We derive closed-form evaluations for sensing loss functions for this specific task, and further provide corresponding numerical results.
	\item In Section \ref{sec:conclusion}, we conclude this paper.
\end{itemize}
Moving beyond \cite{wang_2026_TIT}, we provide a systematic theoretical characterization of Pareto-optimal input distributions under various power constraints, establishing their discrete nature and further revealing the structures of their supports. Importantly, our analytical and numerical results reveal that sensing favors deterministic transmit power, whereas communications benefit from increasingly diverse power levels, thereby unveiling the fundamental role of power randomness in shaping the sensing-communication tradeoff. These findings provide theoretical guidance for the design of practical discrete constellations, such as amplitude and phase-shift keying constellations \cite{shao2026designapskconstellationsapproaching}, to approach the fundamental ISAC Pareto boundary.

\emph{Notations:} Upright symbols denote random variables, as $\mathsf{a}$, $\boldsymbol{\mathsf{a}}$, and $\boldsymbol{\mathsf{A}}$ respectively represent a random scalar, column vector, and matrix; while their deterministic realizations are italic, e.g., $a$, $\boldsymbol{a}$, and $\boldsymbol{A}$.
$\mathbb{R}$, $\mathbb{C}$, $\mathbb{N}$, and $\mathbb{S}$ respectively represent the set of real numbers, complex numbers, positive integers, and unit sphere. We use $\mathbb{P}_{\mathsf{a}}$ to denote the probability measure of the random variable $\mathsf{a}$. $\jmath$ is the imaginary unit such that $\jmath^2=-1$, and $\mathrm{e}\approx 2.718$ is the base of the natural logarithm. In addition, $\I_N$ denotes the $N\times N$ identity matrix; $\boldsymbol{0}_{M\times N}$ denotes the $M\times N$ all-zero matrix matrix. The dimensions and subscripts are omitted wherever there is no confusion. Finally, the notations $\mathcal{CN}$, $\mathcal{N}$, and $\mathcal{U}$ respectively denote circularly symmetric complex Gaussian, real Gaussian, and uniform distributions.

\emph{Operators:} $(\cdot)^*$, $(\cdot)^{-1}$, $(\cdot)^\trans$, and $(\cdot)^\herm$ are respectively the complex conjugation, inverse, transpose, and Hermitian transpose operators for matrices. $\tr(\cdot)$ denotes the trace of its matrix argument. $\mathbb{E}[\cdot]$ is the statistical expectation operator. $\vect(\cdot)$ vectorizes a matrix by stacking its columns; $\diag(\cdot)$ forms a diagonal matrix from its arguments. $\supp(\cdot)$ denotes the support of a distribution/probability measure. $\card\{\cdot\}$ denotes the cardinality of the given set. $\Re(\cdot)$ and $\Im(\cdot)$ represent the real and imaginary parts of the argument, respectively. $\{a\}^+ := \max\{0,a\}$.
Finally, throughout this paper, all logarithms are natural, i.e., $\log(\cdot)=\log_{\mathrm{e}}(\cdot)$, and hence all information-related quantities (e.g., achievable rate and mutual information) are measured in $\mathrm{nats}$.

\section{System Setup}
\label{sec:system_model}

\subsection{Measurement Model}
The measurement model in this paper generalizes the single-input single-output
setup considered in \cite{bistatic_ISAC_TIT} to a more general multi-receiving-antenna setup:
\begin{subequations}
	\label{eqn:measurement_model}
	\begin{align}
			\Yc &= \x\hc^\trans + \Zc, \label{eqn:measurement_model_c}\\
		\Ys &= \x\hs^\trans(\boldeta) + \Zs \label{eqn:measurement_model_s},
	\end{align}
\end{subequations}
where 
\begin{itemize}
	\item $\x\in\mathbb{C}^{N}$ is the transmitted (time-domain) ISAC signal;
	\item $\hc\in\mathbb{C}^{\Lhc}$ and $\hs(\boldeta)\in\mathbb{C}^{\Lhs}$ are the $\Lhc$- and $\Lhs$-dimensional communications and sensing channel, respectively.\footnote{For simplicity of subsequent theoretical exposition and notation, the communication channel $\hc$ is assumed to be deterministic. It is straightforward to consider fading channels by taking the expectation on the rate over its distribution to yield ergodic communication performance \cite{wang_2026_TIT}.} In particular, the sensing channel $\hs$ is a function of the $\Ln$-dimensional real-valued unknown parameter $\boldeta\in\mathbb{R}^{\Ln}$; 
	\item $\Zc\in\mathbb{C}^{N\times\Lhc}$ and $\Zs\in\mathbb{C}^{N\times\Lhs}$ are additive Gaussian noise to the communications and sensing channels, respectively, i.e.,
	\begin{eqnarray*}
		\begin{aligned}
			\vect\{\Zc\}&\sim\mathcal{CN}(\mathbf{0}_{\Lhc N},\noisevar\I_{\Lhc N}), \\
			\vect\{\Zs\}&\sim\mathcal{CN}(\mathbf{0}_{\Lhs N},\noisevar\I_{\Lhs N}),
		\end{aligned}
	\end{eqnarray*}
	with $\noisevar$ being the noise variance; and
	\item $\Yc\in\mathbb{C}^{N\times\Lhc}$ and $\Ys\in\mathbb{C}^{N\times\Lhs}$ are the overall observation matrix at the communications and sensing \acp{Rx}, respectively.
\end{itemize}

The goal for the considered ISAC system is twofold. For communications, it aims to convey downlink communication information to the communication \ac{Rx} through channel $\hc$, which is assumed to be known before each transmission \cite{Xiong_TIT,wang_2026_TIT}. The sensing task is to estimate unknown parameter $\boldeta$ (by estimating $\hs$) based on radar-related principles. Specifically, the ISAC waveform $\x$ is regarded as a known reference signal at the sensing \ac{Rx} for parameter sensing.

To investigate the fundamental sensing-communication tradeoff, consider the following three types of power constraint on input waveform $\x$:
\begin{itemize}
	\item[i)] General power constraint \cite{wang_2026_TIT,Xiong_TIT}:
	\begin{equation}
		\mathbb{E}\big[\|\x\|^2\big]\leq N\rho:=B.
		\label{eqn:GP}
	\end{equation}
	\item[ii)] Maximum amplitude constraint \cite{SMITH1971}:
	\begin{equation}
		\|\x\|\leq \sqrt{N}\Ama := A.
		\label{eqn:MA}
	\end{equation}
	\item[iii)] Constant modulus constraint:
	\begin{equation}
		\|\x\|= \sqrt{N}\Acm.
		\label{eqn:CM}
	\end{equation}
\end{itemize}

\subsection{ISAC Performance Metric}
\subsubsection{Communications Metric}
The communication performance is evaluated by the achievable data rate \cite{wang_2026_TIT,Xiong_TIT}, i.e., 
\begin{equation}
	\Rate[\Px] := \frac1N\MI(\x;\Yc|\hc),
	\label{eqn:R_def}
\end{equation}
where $\MI(\cdot|\hc)$ is the (conditional) mutual information on the channel $\hc$, and $\Px$ is the probability measure of the input ISAC waveform $\x$. When $\hc$ is known at the communication \ac{Rx}, right-multiplying both sides of the measurement model \eqref{eqn:measurement_model_c} with the quantity $\frac{\hc^*}{\sigma_{\mathsf z}\|\hc\|}$ yields the equivalent communication channel
\begin{equation}
	\yc:=\frac{\Yc\hc^*}{\sigma_{\mathsf z}\|\hc\|} = \frac{\|\hc\|}{\sigma_{\mathsf z}}\x+\underbrace{\frac{\Zc\hc^*}{\sigma_{\mathsf z}\|\hc\|} }_{:=\zc} = \sqrt{\SNRc}\x+\zc,
	\label{eqn:y_Yc}
\end{equation}
where $\SNRc:=\|\hc\|^2/\noisevar$ is the effective channel gain. It is clear that $ \zc	\sim\mathcal{CN} (\boldsymbol0,\I_N)$. Since $\yc$ is a sufficient statistic for $\x$
based on $\Yc$, we have 
$$\MI(\x;\Yc|\hc)=\MI(\x;\yc).$$ 
Note that $\MI(\x;\yc)$ also depends on $\hc$, through the scalar $\SNRc$. In subsequent analysis, the mutual information is evaluated using the equivalent channel output $\yc$.

\subsubsection{Sensing Metric}

For the waveform $\x$, the Bayesian-type \ac{CRB} matrix for estimating parameter $\boldeta$ is given as
\begin{equation}
	\CRBm(\x) := \Fr_{\mathrm{Bayesian}}^{-1}(\x),
	\label{eqn:CRB_def}
\end{equation}
where $\Fr_{\mathrm{Bayesian}}$ is the Bayesian \ac{FIM} composed of the averaged conditional \ac{FIM} and the prior \ac{FIM}, i.e.,
$$\Fr_{\mathrm{Bayesian}}(\x)=\Fr(\x)+\Fp.$$
Specifically, the \ac{FIM} $\Fr$ is evaluated as 
\begin{equation*}
	\Fr(\x):= \mathbb{E}\Bigg[\frac{\partial \ln f_{\Ys|\boldeta,\x}(\Ys|\boldeta,\x)}{\partial \boldeta}\frac{\partial \ln f_{\Ys|\boldeta,\x}(\Ys|\boldeta,\x)}{\partial \boldeta^\trans}\Bigg|\x\Bigg],
\end{equation*}
with $f_{\Ys|\boldeta,\x}(\cdot|\cdot)$ denoting the likelihood function of $\Ys$, and $\Fp$ is the \ac{FIM} from the prior knowledge over the distribution of $\boldeta$, given as \cite{Xiong_TIT}
\begin{equation*}
	\Fp:=\mathbb{E}\Bigg[\frac{\partial\ln f_{\boldeta}(\boldeta)}{\partial \boldeta}\frac{\partial\ln f_{\boldeta}(\boldeta)}{\partial \boldeta^\trans}\Bigg],
\end{equation*}
with $f_{\boldeta}(\cdot)$ being the prior distribution of $\boldeta$.

It is well-known that practical estimators (e.g. the maximum a posteriori estimator) can asymptotically achieve the Bayesian-type \ac{CRB} in the high \ac{SNR} regime \cite{Xiong_TIT,wang_holigraphic_TWC}. Hence, in this paper, the expected trace of the weighted \ac{CRB} matrix is defined as the sensing metric, i.e.,
\begin{equation}
	\xi[\Px] := \mathbb{E}\left[\tr\Big\{\W\cdot\CRBm(\x)\Big\}\right],
	\label{eqn:xi_def}
\end{equation}
where $\W\succeq\boldsymbol{0}$ is a diagonal weighting matrix.  The purpose of introducing $\W$ is to cast more interest on some particular parameters in $\boldeta$, or perhaps to compensate for and balance the dimensionality for different parameters in $\boldeta$ \cite{range_compression_TSP,wang2026unifiedcramerraoboundframework} (such as angles versus Doppler).  

It is worthwhile to note that the metric \eqref{eqn:sensing_error} is a Miller-Chang type \ac{CRB} \cite{Miller1978MCRB}. In particular, it treats $\x$ as a nuisance parameter, and takes the expectation with respect to the CRB evaluated at $\x$. This treatment is particularly relevant to the scenario when realization of $\x$ is perfectly known at the sensing \ac{Rx}. Similar metric for sensing has been reported in \cite{Xiong_TIT,wang_2026_TIT}, which can be regarded as the ``ergodic error'' that is averaged over each possible realization of $\x$. Based on the \ac{CRB} \eqref{eqn:CRB_def}, an explicit expression for the defined sensing distortion \eqref{eqn:xi_def} is provided in the following proposition.

\begin{proposition} \label{proposition:closed_CRB}
	The \ac{CRB}-related sensing distortion for estimating $\boldeta$ can be expressed as
	\begin{equation} 
		\xi[\Px]= \mathbb{E}\big[\phi(\|\x\|^2)\big],
		\label{eqn:sensing_error}
	\end{equation}
	where 
	\begin{equation}
		\phi(s) := \sum_{\ell=1}^{\Ln} \frac{\lambda_\ell}{1+\sigma_\ell s}.
		\label{eqn:phi_def}
	\end{equation}
	Here, $\lambda_\ell$ is the $\ell$-th diagonal element of matrix $\boldLambda$, i.e., $\lambda_\ell:=[\boldLambda]_{\ell,\ell}$, where the matrix $\boldLambda$ is defined as
	\begin{equation}
		\boldLambda:=\boldsymbol Q^\trans\Fp^{-1/2}\W\Fp^{-1/2}\boldsymbol Q.
		\label{eqn:Lambda_def}
	\end{equation}
	Here, $\boldsymbol{Q}$ is a real-valued orthogonal matrix from the eigenvalue decomposition of the matrix $\boldsymbol{G}$, i.e., 
	\begin{equation}
		\boldsymbol{G}=\boldsymbol{Q}\boldSigma\boldsymbol{Q}^\trans=\boldsymbol{Q}\diag\{\sigma_1,\cdots,\sigma_{\Ln}\}\boldsymbol{Q}^\trans,
		\label{eqn:G_decomp}
	\end{equation}
	where $\boldsymbol{G}$ is given as
	\begin{equation}
		\boldsymbol G:=
		\frac{2}{\noisevar}\Fp^{-1/2}
		\mathbb{E}\left[
		\Re\left\{
		(\nabla_{\boldeta}\hs)^\herm (\nabla_{\boldeta}\hs)\right\}\right]
		\Fp^{-1/2}.
		\label{eqn:G_def}
	\end{equation}
	Here, $\nabla_{\boldeta}\hs$ is defined as
	\begin{equation*}
		\nabla_{\boldeta}\hs
		:=
		\frac{\partial\hs(\boldeta)}
		{\partial\boldeta^\trans}
		=
		\begin{bmatrix}
			\frac{\partial}{\partial[\boldeta]_1}
			&\cdots&
			\frac{\partial}{\partial[\boldeta]_{\Ln}}
		\end{bmatrix}\hs(\boldeta)\in\mathbb{C}^{\Lhs\times\Ln}.
	\end{equation*}
	Finally, $\sigma_\ell$ is the $\ell$-th eigenvalue of $\boldsymbol{G}$, i.e., $\sigma_\ell=[\boldSigma]_{\ell,\ell}$ in \eqref{eqn:G_decomp}.
	\begin{proof}
		See Appendix \ref{appendix:closed_CRB}.
	\end{proof}
\end{proposition}

\begin{remark}
	Here, $\sigma_\ell$ and $\lambda_\ell$ are referred to as the ``effective sensing information gain'' and ``effective sensing distortion coefficient'' of the $\ell$-th sensing mode, respectively. Specifically, $\sigma_\ell$ is closely related to the Fisher information for parameter estimation, whereas $\lambda_\ell$ quantifies the contribution of the corresponding mode to the weighted sensing distortion. Both coefficients are non-negative. In the sequel, we assume
	\begin{equation}
		\lambda_\ell\sigma_\ell>0, \qquad \exists \ell\in\{1,\cdots,\Ln\}.
		\label{eqn:sensing_condition}
	\end{equation}
	This ensures that the sensing distortion is strictly decreasing with transmit power and that the sensing task exhibits a non-trivial dependence on the transmitted waveform $\x$. In addition, we assume $\Fp$ is positive definite, i.e., $\Fp\succ \boldsymbol{0}$.
\end{remark}
\section{Main Theoretical Results}
\label{sec:results}

In this section, we present our main theoretical results. Before proceeding, we first formulate the following functional Pareto optimization that reflects the CRB-Rate tradeoff \cite{wang_2026_TIT}:
\begin{equation}
	\begin{aligned}
		&\sup_{\Px}& \quad &\alpha \frac{\Rate[\Px]}{\Rate^\star} - (1-\alpha)\frac{\xi[\Px]}{\xi^\star}\\
		&\mathrm{s.t.}& \quad & \mathrm{some\ power\ constraint(s).}
	\end{aligned}
	\label{eqn:Pareto_original}
\end{equation}
Here, $\alpha\in[0,1]$ is the sensing-communication tradeoff preference, i.e.,
small values of \(\alpha\) correspond to a preference for sensing, and large values of \(\alpha\) correspond to a preference for communication. Moreover,
\begin{equation*}
	\Rate^\star:=\max_{\Px}\ \Rate[\Px],\qquad \xi^\star:=\min_{\Px}\ \xi[\Px]
\end{equation*}
are the maximum achievable rate and the minimum achievable distortion, under the same constraints as \eqref{eqn:Pareto_original}, which correspond to communication- and sensing-optimal points (i.e., two end points on the Pareto boundary), respectively. The power constraints for \eqref{eqn:Pareto_original} can be either GP, MA, or CM, corresponding to \eqref{eqn:GP}, \eqref{eqn:MA}, or \eqref{eqn:CM}, respectively. The normalization in \eqref{eqn:Pareto_original} is to ensure dimensional consistency and to guarantee the tradeoff parameter $\alpha$ really reflects the preference between sensing and communication. By sweeping $\alpha$ from $0$ to $1$, we trace out the entire Pareto-boundary from the sensing- to communication-optimal point. 

It should be noted that the optimization is performed over $\Px$ \cite{wang_2026_TIT}. In essence, problem \eqref{eqn:Pareto_original} aims to explore, at each point on the Pareto boundary (corresponding to some $\alpha$), the statistical properties of the optimal input waveform $\x$.

The sensing distortion characterized in Proposition \ref{proposition:closed_CRB} exhibits a distinctive dependence on the transmitted waveform: it is solely determined by the total transmit power, rather than the specific direction (e.g., phase) of the waveform. Specifically, defining the transmit power as 
\begin{equation*}
	\s:=\|\boldsymbol \x\|^2,
\end{equation*}
and denoting its probability measure as $\Ps$, the sensing distortion \eqref{eqn:sensing_error} can be functionally expressed as
\begin{equation}
	\xi[\Ps]=\mathbb{E}\big[\phi(\s)\big]=\int \sum_{\ell=1}^{\Ln}\frac{\lambda_\ell}{1+\sigma_\ell s}\,\diff \Ps(s).
	\label{eqn:xi_ps}
\end{equation}
In contrast, the communication rate \eqref{eqn:R_def} generally depends on both the power/amplitude and the direction distributions of the transmitted waveform. This observation motivates decomposing the input waveform $\x$ into its radial and directional components, namely,
\begin{equation}
	\x=\sqrt{\s}\u,
	\label{eqn:x_decomp}
\end{equation}
where $\u\in\mathbb C^N$ is a unit-norm random directional vector satisfying $\|\u\|^2=1$ almost surely. Under \eqref{eqn:x_decomp}, it is clear that $\Px$ is completely described by $\mathbb{P}_{\s,\u}=\Ps \mathbb{P}_{\u|\s}$; the achievable rate in \eqref{eqn:R_def} can thus be written as a functional of  $\Ps$ and $\mathbb{P}_{\u|\s}$, i.e., $\MI[\Ps,\mathbb{P}_{\u|\s}]$.

\begin{lemma}[Optimal Directional Distribution for the Mutual Information]
	\label{lemma:rate_opt}
	For any feasible $\Ps$ and any conditional directional distribution
	$\mathbb{P}_{\u|\s}$, the following inequality holds:
	\begin{equation}
		\MI[\Ps,\mathbb{P}_{\u|\s}] \leq \MI[\Ps,\mu_N] = \MI(\sqrt{\s}\u^\star;\yc),
		\label{eqn:MI_inequality}
	\end{equation}
	where $\mu_N:=\mathcal{U}(\mathbb{S}_{\mathbb{C}}^{N-1})$ denotes the uniform probability measure on the $N$-dimensional complex unit sphere specified as $\mathbb S_{\mathbb C}^{N-1}
	:=\{\boldsymbol u\in\mathbb C^N:\|\boldsymbol u\|^2=1\}$, and the optimal directional vector $\u^\star\sim\mu_N$ is independent of $\s$.
	\begin{proof}
		The result follows from the unitary invariance of the Gaussian channel and the concavity of mutual information with respect to the input distribution, using the symmetrization argument \cite[Chapter 6.2]{Polyanskiy2025}. Specifically, averaging any input distribution over the unitary group preserves its power distribution while yielding a uniform directional distribution, without decreasing mutual information.
	\end{proof}
\end{lemma}

\begin{remark}
	Since fundamental performance limits are explored, inequality \eqref{eqn:MI_inequality} in Lemma \ref{lemma:rate_opt} can be tightened to equality (by adopting the optimal uniform directional distribution). Hence, subsequent analysis of the mutual information is restricted to $\MI[\Ps,\mu_N]$, which is hereafter abbreviated as $\MI[\Ps]$ for notational convenience, with the implicit understanding that the directional component follows $\u^\star\sim\mu_N$.
\end{remark}

\begin{remark}
	The above results reveal that the sensing and communication objectives exhibit fundamentally different dependencies on the input distribution. Specifically, the sensing distortion depends solely on the power distribution $\Ps$, whereas the communication mutual information can be maximized by a uniform directional distribution $\mu_N$ for each power realization. This reduces the original problem \eqref{eqn:Pareto_original} to an optimization over the scalar power distribution $\Ps$, as stated in the next proposition. This structural characterization provides the foundation for investigating the support properties of Pareto-optimal power distributions in the subsequent analysis. 
\end{remark}

\begin{proposition}
	\label{proposition:power_domain}
	The Pareto optimization problem in \eqref{eqn:Pareto_original} can be equivalently reduced to
	\begin{equation}
		\begin{aligned}
			&\sup_{\Ps}& \quad & \Fobj[\Ps] := \MI[\Ps] + \beta(\alpha){\xi}[\Ps]\\
			&\mathrm{s.t.}& \quad & \mathrm{some\ power\ constraint(s)},
		\end{aligned}
		\label{eqn:Pareto}
	\end{equation}
	where 
	\begin{equation}
		\beta(\alpha):=-\frac{(1-\alpha)N\Rate^\star}{\alpha \xi^\star}\in(-\infty,0],
		\label{eqn:beta_def}
	\end{equation}
	$\MI[\Ps]=\MI(\sqrt{\s}\u^\star;\yc)$ and $\xi[\Ps]$ is defined in \eqref{eqn:xi_ps}.
	\begin{proof}
		The objective in \eqref{eqn:Pareto_original} can be rewritten as 
		\begin{equation*}
			\begin{aligned}
				&\ \sup_{\Px} \quad \alpha \frac{\mathcal{R}[\Px]}{\mathcal{R}^\star} - (1-\alpha)\frac{\xi[\Px]}{\xi^\star} \\
				=&\ \sup_{\mathbb{P}_{\s,\u}}\quad  \frac{\alpha}{N\Rate^\star}\Big( \MI[\Ps,\mathbb{P}_{\u|\s}]  -\frac{(1-\alpha)N\Rate^\star}{\alpha\xi^\star} \xi[\Ps]\Big) \\
				=&\  \frac{\alpha}{N\Rate^\star}\times \sup_{\mathbb{P}_{\s,\u}}\quad  \MI[\Ps,\mathbb{P}_{\u|\s}]  +\beta \xi[\Ps] \\
				=&\ \frac{\alpha}{N\Rate^\star}\times \sup_{\Ps} \quad \Big(\sup_{\mathbb{P}_{\u|\s}}\quad \MI[\Ps,\mathbb{P}_{\u|\s}] + \beta \xi[\Ps]\Big) \\
				=&\  \frac{\alpha}{N\Rate^\star}\times \sup_{\Ps} \quad   \MI[\Ps,\mu_N] + \beta \xi[\Ps],
			\end{aligned}
		\end{equation*}
		where the last equality holds according to Lemma \ref{lemma:rate_opt}. Ignoring the constant scaling $\frac{\alpha}{N\Rate^\star}$ that is independent of the optimization variable $\Ps$ yields the objective in \eqref{eqn:Pareto}.
	\end{proof}
\end{proposition}

Before characterizing the Pareto-optimal power distributions, we first establish the existence of an optimizer for the reduced functional problem \eqref{eqn:Pareto}.

\begin{lemma}[Existence of an Optimal Power Distribution]
	\label{lemma:existance_Psopt}
	
	Under either the GP constraint with $\rho>0$ or the MA constraint with $A>0$, the energy-domain Pareto optimization admits an optimal probability measure $\Psopt$.
	
	\begin{proof}
		Under GP, the average-power constraint ensures uniform tightness of the feasible probability measures, while the moment constraint is preserved under weak convergence. Hence, the feasible set is weakly compact by Prokhorov's theorem \cite[Chapter 1.5]{Prokhorov}. Under MA, weak compactness follows directly from the compactness of the admissible power domain $[0,A^2]$. Since the Pareto objective is upper semicontinuous under weak convergence on either feasible set, it attains its maximum.
	\end{proof}
\end{lemma}

Having established the existence of optimal power distributions, we can replace the $\sup$ operator in the objective with the $\max$ operator in \eqref{eqn:Pareto} hereafter. 	We next examine how the corresponding Pareto-optimal communication and sensing performances vary with the tradeoff preference $\alpha$.

\begin{corollary}
	For $0\leq \alpha_1 < \alpha_2 \leq 1$, denote $\mathbb{P}_{\s,\alpha_1}^\star$ and $\mathbb{P}_{\s,\alpha_2}^\star$ as optimal probability measures to \eqref{eqn:Pareto} corresponding to $\alpha_1$ and $\alpha_2$, respectively. We have
	\begin{equation*}
		\MI[\mathbb{P}_{\s,\alpha_1}^\star] \leq \MI[\mathbb{P}_{\s,\alpha_2}^\star] , \qquad  \xi[\mathbb{P}_{\s,\alpha_1}^\star] \leq \xi[\mathbb{P}_{\s,\alpha_2}^\star].
	\end{equation*}	
	\begin{proof}
		For $0\leq \alpha_1 < \alpha_2 \leq 1$, we have $\beta(\alpha_1) < \beta(\alpha_2)\leq 0$. By Pareto-optimality of $\mathbb{P}_{\s,\alpha_1}^\star$ and $\mathbb{P}_{\s,\alpha_2}^\star$, it immediately follows
		\begin{equation*}
			\begin{aligned}
				\MI[\mathbb{P}_{\s,\alpha_1}^\star] + \beta(\alpha_1) \xi[\mathbb{P}_{\s,\alpha_1}^\star] &\geq \MI[\mathbb{P}_{\s,\alpha_2}^\star] + \beta(\alpha_1) \xi[\mathbb{P}_{\s,\alpha_2}^\star],\\
				\MI[\mathbb{P}_{\s,\alpha_2}^\star] + \beta(\alpha_2) \xi[\mathbb{P}_{\s,\alpha_2}^\star] &\geq \MI[\mathbb{P}_{\s,\alpha_1}^\star] + \beta(\alpha_2) \xi[\mathbb{P}_{\s,\alpha_1}^\star].
			\end{aligned}
		\end{equation*}
		Adding these two inequalities and rearranging terms yield
		\begin{equation*}
			\big(\beta(\alpha_1)-\beta(\alpha_2)\big)\big(\xi[\mathbb{P}_{\s,\alpha_1}^\star]-\xi[\mathbb{P}_{\s,\alpha_2}^\star]\big)  \geq 0
		\end{equation*}
		As $\beta(\alpha_1)<\beta(\alpha_2)$, $\xi[\mathbb{P}_{\s,\alpha_1}^\star]\leq \xi[\mathbb{P}_{\s,\alpha_2}^\star]$. From the second inequality: 
		$$\MI[\mathbb{P}_{\s,\alpha_2}^\star] - \MI[\mathbb{P}_{\s,\alpha_1}^\star]   \geq \beta(\alpha_2) (\xi[\mathbb{P}_{\s,\alpha_1}^\star]-\xi[\mathbb{P}_{\s,\alpha_2}^\star] ) \geq 0,$$ 
		we thus have $\MI[\mathbb{P}_{\s,\alpha_2}^\star] \geq \MI[\mathbb{P}_{\s,\alpha_1}^\star] $.
	\end{proof}
\end{corollary}

From the above corollary, we know that both the Pareto-optimal communication rate and the sensing distortion are non-decreasing with the communication preference $\alpha$. This is intuitive, since if we give more weight to communications, we generally expect larger mutual information and worse sensing accuracy (e.g., larger error). Beyond the monotonicity of the Pareto-optimal performances, we further investigate the continuity of the optimal power distribution itself with respect to $\alpha$.

\begin{corollary}
	Under either the GP or MA constraint, the Pareto-optimal power distribution $\Psopt$ is weakly continuous with respect to the sensing-communication preference $\alpha\in[0,1]$. Specifically, for any sequence $\alpha_n\to\alpha$, we have $\mathbb{P}_{\s,\alpha_n}^\star\Rightarrow \Psopt$.
	
	\begin{proof}
		The feasible set of power distributions is weakly compact under either constraint, and the Pareto objective is jointly continuous in the power distribution and $\alpha$. Moreover, the strict concavity of the weighted sum of mutual information and sensing loss ensures the uniqueness of the optimal power distribution for $\alpha>0$, while the sensing-optimal distribution is unique at $\alpha=0$ (see Lemmas \ref{lemma:corner_points_GP} and \ref{lemma:corner_points_MA}). Therefore, the result follows from Berge's maximum theorem \cite[pp. 163]{Berge1963}.
	\end{proof}
\end{corollary}

\subsection{Results for General Power Constraint \eqref{eqn:GP}}

We first consider the GP constraint. To characterize the Pareto-optimal power distributions, we begin by identifying the communication- and sensing-optimal endpoints, which also determine the normalization factors in \eqref{eqn:Pareto}.

\begin{lemma}[$\Rate^\star$ and $\xi^\star$ under GP]
	\label{lemma:corner_points_GP}
	Under GP, the maximum achievable communication rate is 
	\begin{equation}
		\Rate^\star = \log\left(1+\SNRc\rho\right).
		\label{eqn:rate_opt_GP}
	\end{equation}
	This optimal $\Rate^\star$ is achieved when $\x\sim\mathcal{CN}(\boldsymbol{0},\rho \I_{N})$. The minimal sensing error is
	\begin{equation}
		\xi^\star = \sum_{\ell=1}^{\Ln} \frac{\lambda_\ell}{1+\sigma_\ell N\rho} = \phi(B).
		\label{eqn:E_opt_GP}
	\end{equation}
	It is achieved when $\|\x\|^2=N\rho=B$ almost surely.
	
	\begin{proof}
		The maximum achievable rate is simply the capacity of the channel $\hc$ under Gaussian noise, which is explicitly expressed as \eqref{eqn:rate_opt_GP}. For sensing, consider Jensen's inequality:
		\begin{equation*}
			\xi
			= \mathbb{E}\left[\sum_{\ell=1}^{\Ln} \frac{\lambda_\ell}{1+\sigma_\ell\|\x\|^2}\right]\geq \sum_{\ell=1}^{\Ln} \frac{\lambda_\ell}{1+\sigma_\ell\mathbb{E}[\|\x\|^2]}\geq\phi(B),
		\end{equation*}
		where the equality holds when $\|\x\|^2$ is deterministic (which requires full-power utilization, i.e., $\|\x\|^2=B$). 
	\end{proof}
\end{lemma}

With the two endpoints characterized, we next establish the necessary and sufficient optimality conditions for the power distribution under GP, which form the basis for the subsequent support analysis.

\begin{proposition}[KKT Optimality Conditions under GP]
	\label{proposition:KKT_GP}
	For any $\alpha\in(0,1)$, a feasible power distribution $\Psopt$ is optimal for \eqref{eqn:Pareto} under the GP constraint \eqref{eqn:GP} if and only if there exist constants $\nu_\alpha\geq 0$ and $c_\alpha\in\mathbb R$ such that
	\begin{equation}
		\begin{cases}
			l_\alpha(s;\Psopt) \leq 0,
			& \forall s\geq 0\\[2mm]
			l_\alpha(s;\Psopt) = 0,
			& s\in\operatorname{supp}(\Psopt),
		\end{cases}
		\label{eqn:KKT_GP}
	\end{equation}
	where
	\begin{equation}
		\l_\alpha(s;\Ps):=\kappa(s;\Ps)
		+\beta\phi(s)-\nu_\alpha s - c_\alpha,
		\label{eqn:KKT_density}
	\end{equation}
	together with the complementary slackness condition
	\begin{equation}
		\nu_\alpha
		\left(
		\int_0^\infty s\,\diff \Psopt(s)-N\rho
		\right)=0.
		\label{eqn:KKT_GP_slack}
	\end{equation}
	Here, $\kappa(s;\Ps)$ denotes the marginal entropy density in the power domain, defined as
	\begin{equation}
		\kappa(s;\Ps)
		:=
		-\int
		p_{\yc|\s}(\yr|s)
		\log p_{\yc}(\yr;\Ps)
		\,\diff\yr,
		\label{eqn:kappa_def}
	\end{equation}
	where $p_{\yc}(\yr;\Ps)$ is the output \ac{PDF} induced by any feasible $\Ps$.
	
	\begin{proof}
		See Appendix \ref{appendix:KKT_GP}.
	\end{proof}
\end{proposition}

The associated KKT conditions \eqref{eqn:KKT_GP} enable us to investigate the structure of the Pareto-optimal support. In particular, the analytic properties of the communication and sensing terms lead to the following discreteness result.

\begin{theorem}[Discreteness of Pareto-Optimal Power Distributions under GP]
	\label{theorem:discreteness_GP}
	
	For $\alpha\in[0,1)$, any optimal power distribution $\Psopt$ satisfying the KKT conditions \eqref{eqn:KKT_GP} in Proposition \ref{proposition:KKT_GP} is purely discrete with locally finite support, i.e., $\supp(\Psopt)=\{s_{\alpha,k}^\star\}_{k\in\mathcal{K}_\alpha}$ or 
	\begin{equation}
		\Psopt
		=
		\sum_{k\in \mathcal K_\alpha}
		p_{\alpha,k}^\star\delta_{s_{\alpha,k}^\star},
		\label{eqn:GP_discrete}
	\end{equation}
	where $\mathcal K_\alpha$ is the finite or countably infinite index set of mass points, 	$p_{\alpha,k}^\star>0$, and $\sum_{k\in\mathcal K_\alpha}p_{\alpha,k}^\star=1$.  More specifically, $\mathcal K_\alpha=\{1,\ldots,K_\alpha\}$ if $K_\alpha:=\card\{\mathcal{K}_\alpha\}<\infty$, and $\mathcal K_\alpha=\mathbb N$ (i.e., $K_\alpha=\infty$) otherwise.
Furthermore,
	\begin{equation}
		\card\Big\{
		\supp(\Psopt)
		\cap[0,M]
		\Big\}<\infty,
		\qquad \forall M\in(0,\infty).
		\label{eqn:GP_local_finite}
	\end{equation}
	Consequently, if the optimal distribution $\Psopt$ has infinitely many mass points, their locations must diverge to infinity.
	
	\begin{proof}
		(Sketch) Using an analytic-function argument inspired by Smith \cite{SMITH1971}, we exploit the distinct analytic properties of the communication and sensing terms. Specifically, suppose that the optimal support has a finite accumulation point. By the identity theorem \cite{Knopp1945}, the KKT equality \eqref{eqn:KKT_GP} must hold identically, contradicting the entire analyticity of the communication term $\kappa(\cdot)$ and the non-removable poles of the sensing term $\phi(\cdot)$. Hence, the optimal power distribution is purely discrete with locally finite support. For details, please see Appendix \ref{appendix:discreteness_GP}. 
	\end{proof}
\end{theorem}

\begin{remark}
	Theorem \ref{theorem:discreteness_GP} establishes that the Pareto-optimal power distribution under GP is purely discrete with no finite accumulation points. Nevertheless, since the admissible power domain is unbounded, this result does not imply a finite number of mass points. In particular, the optimal distribution may possess either finitely or countably infinitely many mass points, with the latter necessarily extending to arbitrarily large power levels.
\end{remark}

Although Theorem \ref{theorem:discreteness_GP} establishes discreteness, it does not determine whether the optimal support is finite or infinite. We next characterize how its cardinality depends on the tradeoff preference $\alpha$.

\begin{theorem}[Support Cardinality under GP]
	\label{theorem:K_alpha_GP}
	Recall $B=N\rho$ and define two preference thresholds
	\begin{equation*}
		\alphas :=
		\frac{N\Rate^\star}{N\Rate^\star+\xi^\star T(B)},
		\qquad
		\alphac :=
		\frac{N\Rate^\star}{N\Rate^\star+\xi^\star D_{\text c}},
		\label{eqn:GP_thresholds}
	\end{equation*}
	where
	\begin{equation*}
		\begin{aligned}
			&T(B) :=
			\sup_{\substack{s\geq0\\s\neq B}}
			\frac{D_\kappa(s)}{D_\phi(s)}, \\
			&D_{\text c} :=
			\frac{N\SNRc^2\rho^2}
			{(1+\SNRc\rho)\left(\sup_{s\geq 0}\{-s\phi'(s)\}\right)},
		\end{aligned}
	\end{equation*}
	with
	\begin{equation}
		\begin{aligned}
			D_\kappa(s)
			&:=\kappa(s;\delta_B)-\kappa(B;\delta_B)
			-\kappa'(B;\delta_B)(s-B),\\
			D_\phi(s)
			&:=\phi(s)-\phi(B)-\phi'(B)(s-B).
		\end{aligned}
		\label{eqn:Dk_Dphi_def}
	\end{equation}
	Here, $\delta_B$ denotes the Dirac probability measure at $B$, and $\kappa'(B;\delta_B)$ denotes the derivative with respect to the first argument, with $\delta_B$ fixed, i.e.,
	\begin{equation*}
		\kappa'(B;\delta_B) = \left. \frac{\partial}{\partial s} \kappa(s;\delta_B)\right|_{s=B}.
	\end{equation*}
	
	Then, $0<\alphas < \alphac<1$, and the following properties hold:
	\begin{subequations}
		\begin{align}
			&\alpha\in[0,\alphas] \Longleftrightarrow \Psopt=\delta_B \Longleftrightarrow K_\alpha=1; \label{eqn:alpha_s}\\
			&\alpha\in(\alphac,1) \Longrightarrow K_\alpha=\infty. \label{eqn:alpha_c}
		\end{align}
	\end{subequations}
	
	\begin{proof}
		See Appendix \ref{appendix:K_alpha_GP}.
	\end{proof}
\end{theorem}

\begin{remark}
	Theorem \ref{theorem:K_alpha_GP} characterizes two distinct support regimes along the communication-sensing Pareto boundary under GP. Specifically, $\alphas$ is the exact threshold for single-mass-point optimality: the sensing-optimal distribution $\delta_B$ remains optimal if and only if $\alpha\in[0,\alphas]$. In contrast, $\alphac$ provides a sufficient threshold for infinitely many mass points, i.e., every optimal power distribution has countably infinite support for $\alpha\in(\alphac,1)$, despite not being communication-optimal. The support cardinality in the intermediate region $\alpha\in(\alphas,\alphac]$ remains analytically uncharacterized and may be either finite or infinite. Nevertheless, Theorem \ref{theorem:K_alpha_GP} guarantees the existence of a non-empty preference interval over which Pareto-optimal power distributions necessarily possess infinitely many discrete mass points. At $\alpha=1$, the communication-optimal input is Gaussian, corresponding to a continuous power distribution.
\end{remark}

Since \(\alphac\) provides a sufficient threshold for infinite-support optimality, its dependence on system parameters reveals how communication and sensing characteristics affect the preference region in which infinite support is guaranteed. Such effects are investigated in the following corollary.

\begin{corollary}[Parametric Properties of $\alphac$ under GP]
	\label{cor:GP_alphac}
	Consider the sufficient threshold $\alphac$ defined in Theorem \ref{theorem:K_alpha_GP}. The following properties hold:
	\begin{enumerate}
		\item \emph{Communication channel gain:}
		$\alphac$ is decreasing in $\SNRc$.
		
		\item \emph{Input dimension:}
		For fixed per-dimensional power $\rho$, $\alphac$ is  increasing in $N$.
		
		\item \emph{Sensing distortion coefficients:}
		$\alphac$ is invariant under a common positive scaling of all $\lambda_\ell$.
		
		\item \emph{Sensing information gains:}
		$\alphac$ is increasing under a common positive scaling of all $\sigma_\ell$.
		
		\item \emph{Transmit power:}
		If $\sigma_\ell>0$ for every $\ell$
		with $\lambda_\ell>0$, then
		\begin{equation*}
			\lim_{\rho\to0^+}\alphac
			=
			\lim_{\rho\to\infty}\alphac
			=1.
		\end{equation*}
		Consequently, $\alphac$ is non-monotonic in $\rho$ and attains at least one global minimum at a finite positive power level.
	\end{enumerate}
	
	\begin{proof}
		See Appendix \ref{appendix:GP_alpha_c}.
	\end{proof}
\end{corollary}

\begin{remark}
	Corollary \ref{cor:GP_alphac} provides further insights into the interplay between communication and sensing requirements in shaping the optimal power distribution. Specifically, a stronger communication channel (e.g., with larger $\SNRc$) enlarges the preference region $(\alphac,1)$ in which infinitely many mass points are guaranteed, whereas a common increase in the effective sensing-information gains (e.g., larger $\{\sigma_\ell\}_{\ell=1}^{\Ln}$) shrinks this region. Interestingly, increasing the input dimension $N$ while keeping the per-dimensional transmit power $\rho$ fixed leaves the maximum achievable communication rate $\Rate^\star$ unchanged, but reduces the minimum sensing distortion $\xi^\star$ due to the increased total transmit power $B=N\rho$, thereby shrinking the guaranteed infinite-support region. These observations are consistent with the distinct roles of power randomness in the two objectives: communication benefits from a richer input distribution, while sensing favors deterministic transmit power, this can also  be interpreted in light of the waveform uncertainty (random) tradeoff established in  \cite{wang_2026_TIT,Xiong_TIT}.\footnote{It should be noted that since $\alphac$ is only a sufficient threshold, these monotonicity properties characterize the guaranteed infinite-support region rather than the exact transition in support cardinality.}
\end{remark}
\subsection{Results for Maximum Amplitude Constraint \eqref{eqn:MA}}
In what follows, we consider the MA constraint, which restricts the transmit power to a bounded interval, in contrast to GP. We begin by characterizing the two endpoints on the Pareto-boundary under this constraint.

\begin{lemma}[$\Rate^\star$ and $\xi^\star$ under MA]
	\label{lemma:corner_points_MA}
	Under MA, the maximum achievable communication rate is 
	\begin{equation}
		\Rate^\star = \frac1N \max_{\substack{\Px\\ \|\x\|\leq A}} \MI(\x;\yc|\hc) = \frac1N \max_{\substack{\Ps\\\supp(\Ps)\subseteq[0,A^2]}} \MI[\Ps].
		\label{eqn:capacity_MA}
	\end{equation}
	This is the classical amplitude-constrained Gaussian channel capacity problem, for which no general closed-form expression is known unfortunately. The capacity-achieving radial (e.g., power) distribution is known to have finitely many mass points, as established in related Gaussian channel settings by Smith~\cite{SMITH1971}, Shamai and Bar-David~\cite{ShamaiBarDavid1989}, and subsequent multidimensional extensions in \cite{Dytso2020,favano_2021_capacity}. Consequently, the capacity and its achieving distribution generally require numerical optimization. The minimal sensing error is
	\begin{equation*}
		\xi^\star = \sum_{\ell=1}^{\Ln} \frac{\lambda_\ell}{1+\sigma_\ell N\Ama^2} = \phi(A^2).
	\end{equation*}
	It is achieved when $\|\x\|=\sqrt{N}\Ama=A$ almost surely. The proof follows the same steps as those in Lemma \ref{lemma:corner_points_GP}.
\end{lemma}

We next establish the KKT optimality conditions under MA. Unlike GP, no average-power multiplier is required for MA, since the power constraint is imposed directly on the admissible domain.

\begin{proposition}[KKT Optimality Conditions under MA]
	\label{proposition:MA_KKT}
	For any $\alpha\in[0,1]$, a feasible power distribution $\Psopt$ is optimal for \eqref{eqn:Pareto}
	under the MA constraint \eqref{eqn:MA} if and only if there exists a constant $c_\alpha\in\mathbb R$ such that
	\begin{equation}
		\begin{cases}
			l_\alpha(s;\Psopt) \leq 0,
			& \forall s\in[0,A^2]\\[2mm]
			l_\alpha(s;\Psopt) = 0,
			& s\in\supp(\Psopt),
		\end{cases}
			\label{eqn:MA_KKT}
	\end{equation}
	where
	\begin{equation}
		\l_\alpha(s;\Ps):=\kappa(s;\Ps)
		+\beta\phi(s) - c_\alpha.
		\label{eqn:KKT_density_MA}
	\end{equation}
	
	\begin{proof}
		The proof follows the same functional variational argument as in the proof of Proposition~\ref{proposition:KKT_GP}. Unlike GP, the MA constraint restricts the admissible power domain to $[0,A^2]$ without imposing an additional average power constraint. Consequently, the corresponding Lagrange multiplier $\nu_\alpha$ in \eqref{eqn:KKT_GP} vanishes, and the KKT conditions reduce to \eqref{eqn:MA_KKT}.
	\end{proof}
\end{proposition}

Combining the KKT conditions with the bounded admissible power domain, we may obtain a stronger support characterization than under GP, as stated in the following theorem.

\begin{theorem}[Finite Discreteness of Pareto-Optimal Power Distributions under MA]
	\label{theorem:MA_discrete}
	
	For any $\alpha\in[0,1]$, every optimal power distribution $\Psopt$ under the MA constraint \eqref{eqn:MA} is finitely discrete, i.e.,
	\begin{equation}
		\Psopt
		=
		\sum_{k\in \mathcal K_\alpha}
		p_{\alpha,k}^\star\delta_{s_{\alpha,k}^\star},
		\qquad K_\alpha:=\card\{\mathcal{K}_\alpha\}<\infty,
		\label{eqn:MA_discrete}
	\end{equation}
	where
	\begin{equation*}
		0\leq s_{\alpha,k}^\star\leq A^2,
		\quad
		p_{\alpha,k}^\star>0,
		\quad
		\sum_{k=1}^{K_\alpha}p_{\alpha,k}^\star=1.
	\end{equation*}
	
	\begin{proof}
		See Appendix \ref{appendix:MA_discrete}.
	\end{proof}
\end{theorem}

\begin{remark}
	Unlike GP, under which the Pareto-optimal power distribution must possess countably infinitely many mass points due to the unbounded power domain for $\alpha\in(\alphac,1]$, MA guarantees a finite number of mass points for all tradeoff preferences, including the communication-optimal endpoint. This distinction follows from the bounded power domain under MA, which, together with the absence of finite accumulation points established by the analyticity argument, ensures finite support.
\end{remark}

 Beyond finite discreteness, the MA constraint allows further characterization of the optimal support, including the presence of the maximum admissible power and the preference thresholds governing single-mass-point optimality and zero-power exclusion.

\begin{theorem}
	\label{theorem:MA_Ps}
	Under MA, the Pareto-optimal probability measures have the following properties:
	\begin{enumerate}
		\item For all $\alpha\in[0,1]$, $A^2\in\supp(\Psopt)$.
		\item \textit{Single-mass-point region}. In this region, we have
		\begin{equation}
			\alpha\in[0,\alphas] \Longleftrightarrow K_\alpha = 1 \Longleftrightarrow \Psopt=\delta_{A^2}.
			\label{eqn:single_mass_region_MA}
		\end{equation}
		Here, 
		\begin{equation}
			\alphas  := \frac{N\Rate^\star}{N\Rate^\star+\xi^\star T(A^2)},  
			\label{eqn:alpha_s_MA} 
		\end{equation}
		where 
		\begin{equation}
			T(A^2) := \left \{
			\sup_{\substack{0\leq s\leq A^2}}
			\frac{\kappa(s;\delta_{A^2})-\kappa(A^2;\delta_{A^2})}{\phi(s)-\phi(A^2)}\right\}^+ .
			\label{eqn:T_def_MA}
		\end{equation}
		\item For all $\alpha\in[0,\alpha_0)$, $0\notin\supp(\Psopt)$, where 
		\begin{equation*}
			\alpha_0
			:=
			\frac{N\Rate^\star}
			{N\Rate^\star+\xi^\star \frac{\SNRc\left\{\frac{\SNRc A^2}{N}-1\right\}^+}{\sum_{\ell=1}^{\Ln}\lambda_\ell\sigma_\ell}}.
		\end{equation*}
	\end{enumerate}
	\begin{proof}
		See Appendix \ref{appendix:MA_Ps}.
	\end{proof}
\end{theorem}

\begin{remark}
	Theorem \ref{theorem:MA_Ps} reveals that the maximum admissible power always belongs to the optimal support under MA, while the sensing-optimal single-mass-point distribution remains optimal for all $\alpha\leq\alphas$. As $\alpha$ increases beyond this threshold, additional mass points necessarily emerge to improve communication performance. Notably, unlike conventional amplitude-constrained capacity maximization, where zero-power signaling may be employed to convey information (e.g., on-off signaling) \cite{Dytso2020,Sharma2010}, Theorem \ref{theorem:MA_Ps} further establishes that zero power is strictly excluded from the optimal support for $\alpha<\alpha_0$. This reflects the sensing requirement, since zero-power transmission contributes no waveform-dependent Fisher information and thus incurs a sensing penalty that outweighs its potential communication benefit in this preference regime.
\end{remark}

While Theorem \ref{theorem:MA_Ps} characterizes several structural properties of the optimal support, it does not quantify the number of mass points beyond the single-mass-point regime. We therefore derive preference-dependent bounds on the support cardinality.

\begin{theorem}[Bounds on the Number of Mass Points under MA]
	\label{thm:MA_cardinality_bound}
	For any $\alpha\in(0,1)$, let $K_\alpha$ denote the number of mass points of the Pareto-optimal power distribution $\Psopt$ under MA. Then, we have
	\begin{equation*}
		\Kl\leq K_\alpha\leq\Ku.
		\label{eqn:MA_cardinality}
	\end{equation*}
	Specifically, $\Kl$ and $\Ku$ are given in \eqref{eqn:MA_bounds}. The functions and constants $a(\cdot)$, $b(\cdot)$, $d_j$, $z_j$, and $\Lhseff$ depend only on the system parameters and are explicitly specified in Appendix~\ref{appendix:MA_cardinality_bound}. 
	
	Moreover, both $\Kl$ and $\Ku$ are non-decreasing in $\alpha$.	

	\begin{proof}
		See Appendix~\ref{appendix:MA_cardinality_bound}.
	\end{proof}
\end{theorem}

\begin{figure*}[t]
	\begin{subequations}
		\label{eqn:MA_bounds}
		\begin{align}
			\Kl
			&= \begin{cases}
				1, & \alpha\in[0,\alphas] \\[2mm]
				\max\left\{2, 	\left\lceil
				\exp\{N\Rate^\star-\MI[\delta_{A^2}]+\beta(\xi[P_{\s,1}^\star]-\xi^\star)\}
				\right\rceil
				\right\}, & \alpha\in(\alphas,1]
			\end{cases}
			\label{eqn:MA_Kl}\\
			\Ku
			&= \begin{cases}
				1, & \alpha\in[0,\alphas] \\[2mm]
				1+
				\left\lfloor
				\displaystyle\min_{1\leq j\leq \Lhseff} \inf_{r>|z_j|+A^2}
				\frac{\log\frac{a(r+|z_j|)+|\beta|b(r+|z_j|)}{|\beta|d_j}}
				{\log\frac{r}{|z_j|+A^2}}
				\right\rfloor, & \alpha\in(\alphas,1]
			\end{cases}
			\label{eqn:MA_Ku}
		\end{align}
	\end{subequations}
	\rule{\linewidth}{1pt}
\end{figure*}

\begin{remark}
	The bounds in Theorem \ref{thm:MA_cardinality_bound} may be conservative, particularly the upper bound, due to the successive inequalities employed in its derivation. Nevertheless, such bounds are commonly used in the information-theoretic literature on amplitude-constrained Gaussian channels, where the primary objective is to characterize scaling laws and parameter-dependent trends rather than obtain tight numerical estimates of the support cardinality (see, e.g., \cite{Dytso2020,Wang2026ImprovedLowerBounds}). In particular, Theorem \ref{thm:MA_cardinality_bound} establishes the non-decreasing dependence of both bounds on the communication preference $\alpha$.
\end{remark}
\subsection{Results for Constant Modulus Constraint}

Under the CM constraint \eqref{eqn:CM}, the transmit power is fixed as $\s=N\Acm^2$ almost surely. Consequently, the power distribution is uniquely determined, and the Pareto optimization admits the following simple characterization.

\begin{corollary}
	\label{corollary:CM}
	Under CM, the Pareto-optimal power distribution is
	\begin{equation*}
		\Psopt=\delta_{N\Acm^2},
	\end{equation*}
	for every $\alpha\in[0,1]$. The corresponding communication rate and sensing distortion are
	\begin{equation*}
		\begin{aligned}
			\Rate = \Rate^\star
			&=
			\frac{1}{N}
			\MI(
			\sqrt{N}\Acm\u^\star;
			\yc),\\
			\xi=\xi^\star
			&=
			\phi(NA_{\mathrm{CM}}^2).
		\end{aligned}
	\end{equation*}
\end{corollary}

\begin{remark}
	Unlike GP and MA, the CM constraint admits no non-trivial communication-sensing tradeoff in the considered model. Specifically, the sensing distortion is independent of the directional distribution, while the communication rate is maximized by the uniform directional distribution. Therefore, the achievable performance region has a single Pareto-optimal point, rather than a non-trivial Pareto boundary. The optimal power distribution remains a single Dirac measure for all tradeoff preferences, and no further optimization of the power distribution is required.
\end{remark}

\subsection{Numerical Algorithms}
\subsubsection{Numerical Algorithm under MA Constraint}
\label{subsec:MA_algorithm}

\begin{algorithm}[t]
	\caption{Adaptive BA Algorithm for MA-Constrained
		Pareto Optimization}
	\label{alg:MA_solver}
	\begin{algorithmic}[1]
		\REQUIRE $N$, $\SNRc$, $A$, $\alpha$, $\phi(\cdot)$,
		tolerances $\epsilon_{\mathrm{in}},\ \epsilon_{\mathrm{out}}$
		\STATE Initialize a feasible finite-support
		distribution $\Ps$
		\REPEAT
		\REPEAT
		\STATE Update mass-point probabilities
		using \eqref{eq:MA_probability_update}
		\STATE Optimize mass-point locations
		according to \eqref{eq:MA_mass_update}
		\UNTIL{the inner convergence criterion $\epsilon_{\mathrm{in}}$ is met}
		\STATE Evaluate the optimality gap
		$\Delta_\alpha[\Ps]$ using
		\eqref{eq:MA_KKT_gap}
		\IF{$\Delta_\alpha[\Ps]>\epsilon_{\mathrm{out}}$}
		\STATE Add a mass point at
		$s_{\mathrm{new}}$ given by
		\eqref{eq:app_support_insertion}
		\STATE Initialize its probability and
		reoptimize the enlarged distribution
		\ENDIF
		\UNTIL{$\Delta_\alpha[\Ps]\leq\epsilon_{\mathrm{out}}$}
		\STATE $\Psopt\longleftarrow \Ps$ (``optimal'' in the numerical sense)
		\RETURN $\Psopt$, $\MI[\Psopt]$, $\xi[\Psopt]$
	\end{algorithmic}
\end{algorithm}

Existing numerical approaches, such as the \ac{BA} algorithm employed in \cite{wang_2026_TIT}, directly optimize the distribution of the input waveform $\x$. However, when $\x$ is vector-valued, as in the considered setup, or matrix-valued, the computational complexity of such approaches generally grows exponentially with the input dimension, rendering them computationally intractable. In contrast, the reduction to the scalar power domain enables numerical algorithms originally developed for classical amplitude-constrained Gaussian channels to be adapted to the considered ISAC setting, thereby substantially reducing the dimensionality for optimization.

Motivated by the dynamic-assignment \ac{BA} approach in \cite{Xiao2020DAB,favano_2021_capacity}, we develop a sensing-aware adaptive BA algorithm to compute the Pareto-optimal power distributions numerically under the MA constraint. For a given $\beta\leq 0$, the algorithm maximizes $\MI[\Ps]+\beta\xi[\Ps]$ over probability measures supported on $[0,A^2]$. Specifically, the probabilities and locations of a finite set of mass points are alternately optimized through a sensing-aware BA iteration and constrained optimization, respectively. The support is then adaptively refined based on the KKT optimality conditions in Proposition \ref{proposition:MA_KKT}, with a KKT-based optimality gap used to assess global objective suboptimality.

The algorithm applies to the communication-optimal endpoint by setting $\beta=0$, yielding $\Rate^\star$. The corresponding $\Rate^\star$, together with $\xi^\star=\phi(A^2)$, determines $\beta(\alpha)$ in \eqref{eqn:beta_def}. The Pareto boundary is subsequently computed by varying $\alpha\in(0,1)$. The sensing-optimal endpoint $\alpha=0$ is given analytically by $\Psopt=\delta_{A^2}$. The complete procedure is summarized in Algorithm~\ref{alg:MA_solver}, while the detailed derivations of the numerical quantities, update rules, and optimality-gap expressions are provided in Appendix~\ref{appendix:MA_numerical_algorithm}.

\subsubsection{Numerical Algorithm under GP Constraint}

\begin{algorithm}[t]
	\caption{Grid-Based BA Algorithm for GP-Constrained
		Pareto Optimization}
	\label{alg:GP_solver}
	\begin{algorithmic}[1]
		\REQUIRE $N$, $\SNRc$, $\rho$, $\alpha$, $\phi(\cdot)$, $\{s_k\}_{k=1}^{K}$,
		tolerance $\epsilon$
		\STATE Initialize $p_k>0$ with $\sum_{k=1}^{K}p_k=1$ and $\sum_{k=1}^{K}p_ks_k\leq N\rho$
		\REPEAT
		\STATE Evaluate
		$f_\alpha(s_k;\Ps)$ for all $k$ according to \eqref{eqn:ISAC_pareto_utility_density}
		\STATE Determine $\nu_\alpha\geq0$ by bisection
		to satisfy the average-power constraint
		and complementary slackness
		\STATE Update $\{p_k\}_{k=1}^{K}$
		according to \eqref{eqn:GP_probability_update}
		\STATE Evaluate the KKT optimality gap $\Delta_\alpha[\Ps]$ using \eqref{eqn:GP_KKT_gap}
		\UNTIL{$\Delta_\alpha[\Ps] \leq\epsilon$}
		\STATE $\Psopt\longleftarrow \Ps$  (``optimal'' in the numerical sense)
		\RETURN $\Psopt$, $\MI[\Psopt]$, $\xi[\Psopt]$
	\end{algorithmic}
\end{algorithm}

Unlike MA, the admissible power domain under GP is unbounded, and the optimal power distribution may possess countably infinitely many mass points, as established in Theorems \ref{theorem:discreteness_GP} and \ref{theorem:K_alpha_GP}. Hence, we adopt a grid-based BA algorithm instead of explicitly optimizing the number and locations of mass points.

Specifically, the power domain is discretized into a fixed grid $\{s_k\}_{k=1}^{K}$. Both the grid resolution and the power truncation threshold $ \max\{s_1,s_2,\cdots,s_K\} $  should be sufficiently large, so that the finite dimension approximation accurately captures the Pareto-optimal performance.

 The associated probabilities $\{p_k\}_{k=1}^{K}$ are optimized using the constrained BA update derived in Appendix \ref{appendix:GP_numerical_algorithm}, where the Lagrange multiplier $\nu_\alpha$ is determined (e.g., via bisection) to enforce the average-power constraint \eqref{eqn:GP}. Unlike the MA algorithm, the grid points remain fixed throughout each optimization.

The numerical optimality is assessed using the KKT-based optimality gap derived in Appendix \ref{appendix:GP_numerical_algorithm}. The communication- and sensing-optimal endpoints are obtained analytically from Lemma \ref{lemma:corner_points_GP}. The complete procedure is summarized in Algorithm \ref{alg:GP_solver}.

\section{Case Study and Numerical Results on Angle-of-Arrival Estimation as the Sensing Task}
\label{sec:case_study}

\begin{table}[t]
	\centering
	\caption{Default Parameter Settings}
	\label{table:numerical}
	\renewcommand{\arraystretch}{1.3}
	\begin{tabular}{c|c||c|c}
		\hline
		Parameter & Value & Parameter & Value \\
		\hline
		$N$ & $4$ & $\sigma_{\uptheta}^2$ & $0.01\, \mathrm{rad}^2$ \\
		$\Lhs$ & $4$ & $\theta_0$ & $35^\circ$ \\
		$\noisevar$ and $\sigma_{\mathsf h}^2$ & $1$ & $\|\hc\|$ & $\sqrt{2}$  \\
		\hline
	\end{tabular}
\end{table}

\subsection{Scenario Setup}
In this part, we consider a specific sensing task of interest, namely, the \ac{AoA} estimation. Specifically, consider the parameterized single-target \cite{Xiong_TIT} channel model:
\begin{equation*}
	\hs = \mathsf{h}\v(\uptheta),
\end{equation*}
where $\mathsf{h}\sim\mathcal{CN}(0,\sigma_{\mathsf{ h}}^2)$ is the complex-valued gain of the sensing channel, $\v(\uptheta)$ is an \ac{Rx}-side steering vector given as \cite{wang2025device}\footnote{Without loss of generality, it is assumed that the inter-element antenna spacing is half-wavelength \cite{wang2025device}.}
\begin{equation*}
	\v(\uptheta) := \frac{1}{\sqrt{\Lhs}}\Big[1,\mathrm{e}^{\jmath \pi\sin(\uptheta)},\cdots, \mathrm{e}^{\jmath \pi(\Lhs-1)\sin(\uptheta)}\Big]^\trans,
\end{equation*}
and $\uptheta\sim\mathcal{N}(\theta_0,\sigma_{\uptheta}^2)$ is the \ac{AoA}. Here, we must have $\Lhs\geq 2$ for array signal processing. The unknown parameter set is defined as
\begin{equation*}
	\boldeta:=\big[\uptheta,\Re\{\mathsf{h}\},\Im\{\mathsf{h}\}\big]^\trans.
\end{equation*}
Here, $\uptheta$ is of interest while $\Re\{\mathsf{h}\}$ and $\Im\{\mathsf{h}\}$ are of nuisance. Hence, we can choose a special weighting matrix, e.g.,
\begin{equation*}
	\W := \begin{bmatrix}
		1&0&0\\
		0&0&0\\
		0&0&0\\
	\end{bmatrix},
\end{equation*}
such that $\xi$ in \eqref{eqn:sensing_error} reflects the estimation performance only related to $\uptheta$ in $\boldeta$.\footnote{This special choice of $\W$ is equivalent to computing the CRB for $\uptheta$ using the equivalent \ac{FIM}, obtained by eliminating the nuisance parameters via the Schur complement \cite{wang2025device}.}

\begin{proposition}
	\label{proposition:AOA_est}
	Based on the above configuration, we have the following closed-form evaluations for the \ac{CRB}-related parameters:
	\begin{equation*}
		\begin{aligned}
			\sigma_1 &= \frac{\sigma_\uptheta^2\sigma_{\mathsf h}^2\pi^2
				(\Lhs-1)(2\Lhs-1)}{6\noisevar} \big(1+\mathrm{e}^{-2\sigma_{\uptheta}^2}\cos(2\theta_0)\big), \\
			\sigma_2 &= \sigma_3 = \frac{\sigma_{\mathsf h}^2}{\noisevar}, \\
			\lambda_1 &= \sigma_{\uptheta}^2,\\
			\lambda_2 &= \lambda_3 = 0.
		\end{aligned}
	\end{equation*}
	
	\begin{IEEEproof}
		See Appendix \ref{appendix:AOA_est}.
	\end{IEEEproof}
\end{proposition}

\begin{remark}
	From Proposition \ref{proposition:AOA_est}, the sensing loss function in \eqref{eqn:phi_def} can be expressed as
	\begin{equation*}
		\phi(s) = \frac{1}{\frac1{\sigma_\uptheta^2}+\frac{\pi^2}{6}
				(\Lhs-1)(2\Lhs-1) \Big(1+\frac{\cos(2\theta_0)}{\mathrm{e}^{2\sigma_{\uptheta}^2}}\Big)\sigma_{\mathsf h}^2\,\frac{s}{\noisevar}}.
	\end{equation*}
	It is observed that for this \ac{AoA} estimation task, the sensing loss function involves only a single mode.
	The CRB decreases with the transmit \ac{SNR} $s/\noisevar$, the number of receive antennas $\Lhs$ (it decreases quadratically with an asymptotic scaling of $O(1/\Lhs^{2})$ in the large-array regime), and the average channel power gain $\sigma_{\mathsf{ h}}^2=\mathbb{E}[|\mathsf{h}|^2]$. A smaller AoA prior variance $\sigma_{\uptheta}^2$ generally improves estimation accuracy through stronger prior information, although its effect also depends on the mean AoA $\theta_0$ through the angular sensitivity term. Moreover, the CRB increases with the absolute mean AoA $|\theta_0|$ over $[-\pi/2,\pi/2]$, indicating improved angular estimation accuracy near the array broadside direction compared with the endfire directions.
\end{remark}

\begin{figure*}[t]
	\centering
	\includegraphics[scale=1]{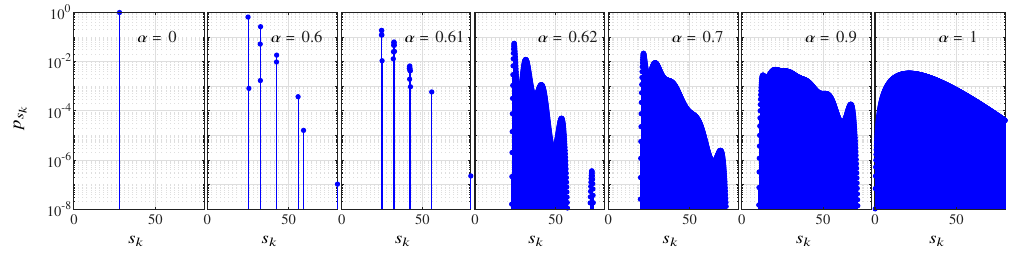}
	\caption{$p_{s_k}$ for various $\alpha$ with $\rho=7$.}
	\label{fig:PMF_GP}
\end{figure*}

\subsection{Numerical Results}
We now provide the numerical results for the single-target \ac{AoA} estimation task discussed in previous subsection. The default configurations that are identical across numerical examples are summarized in Table \ref{table:numerical}. All numerical results in this section are obtained using MATLAB.

\begin{figure}[t]
	\centering
	\includegraphics[scale=1]{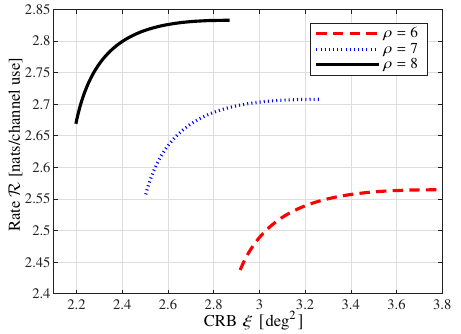}
	\caption{Pareto fronts for the CRB-Rate region for various $\rho$.}
	\label{fig:Pareto_GP}
\end{figure}

\subsubsection{Results for GP} Consider the communication-optimal point, where $\x\sim\mathcal{CN}(\boldsymbol{0},\rho\I_N)$. Therefore, $\s\sim\operatorname{Gamma}(N,\rho)$. The mean and standard deviation for $\s$ are $N\rho$ and $\sqrt{N}\rho$, respectively. To guarantee satisfactory approximation using discrete grid, the power domain is truncated to $[0,N\rho+4\sqrt{N}\rho]$ and uniformly discretized into $K=800$ grid points. The optimization in Algorithm \ref{alg:GP_solver} is terminated when the grid-based KKT optimality gap falls below $\epsilon=10^{-4}$. 

For $\rho=7$, we first calculate that $\alphas=0.167$ and obtain the optimized $p_{s_k}$ (in log-scale) for various tradeoff preference $\alpha$ in Fig. \ref{fig:PMF_GP}. As expected, the sensing-optimal distribution at $\alpha=0$ (in fact, at $\alpha\in[0,\alphas]$) consists of a single mass point at $s=N\rho$. As $\alpha$ increases, additional mass points emerge and the probability mass becomes progressively more dispersed over the power domain. In particular, the distributions exhibit a sparse structure for some $\alpha$, e.g., $\alpha<0.61$, whereas a larger number of grid points acquire non-negligible probabilities as $\alpha$ approaches unity. Moreover, we calculate $\alphac=0.671$, beyond which the optimal power distribution is theoretically guaranteed to possess countably infinitely many mass points, as established in Theorem \ref{theorem:K_alpha_GP}. This is consistent with the increasingly dense support observed numerically for $\alpha>\alphac$. At $\alpha=1$, the optimal power distribution reduces to the Gamma distribution induced by the capacity-achieving Gaussian input. Reviewing the pattern in this figure, we observe a smooth transition from a single mass point, to multiple mass points, and eventually to a continuous power distribution at the communication-optimal point.

By plotting the CRB-Rate pair for each $\alpha\in[\alphas,1]$ in a coordinate system, we observe the Pareto front for the ISAC system, as portrayed in Fig. \ref{fig:Pareto_GP}. In particular, we consider multiple choices of the power constraint $\rho\in\{6,7,8\}$. As expected, both the CRB-Rate region enclosed by the Pareto front expends for larger power budget.

\subsubsection{Results for MA} 

\begin{figure*}[t]
	\centering
	\includegraphics[scale=1]{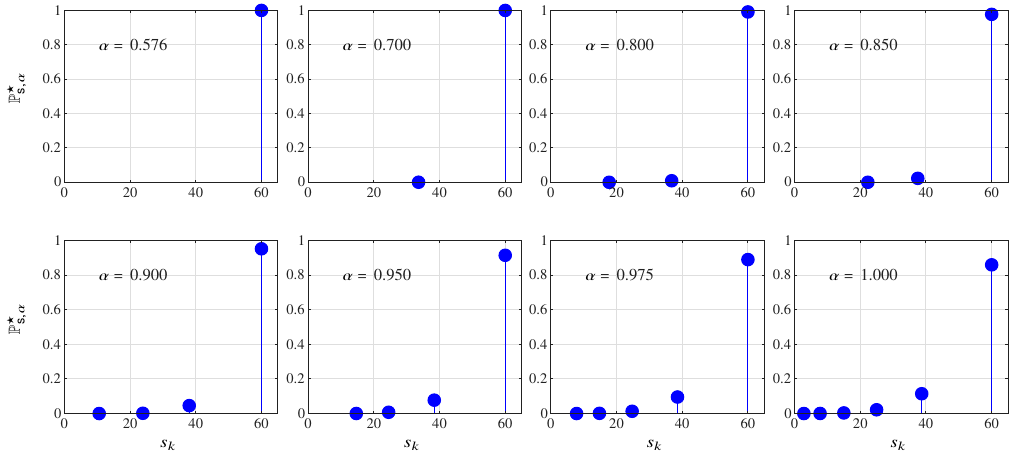}
	\caption{$\Psopt$ for various $\alpha$ with $\Ama=\sqrt{15}$.}
	\label{fig:PMF_MA}
\end{figure*}

\begin{figure}[t]
	\centering
	\includegraphics[scale=1]{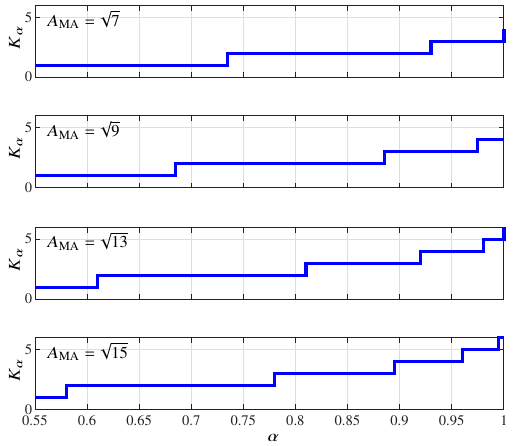}
	\caption{$K_\alpha$ versus $\alpha$ for various $\Ama$.}
	\label{fig:mass_points_MA}
\end{figure}

\begin{figure}[t]
	\centering
	\includegraphics[scale=1]{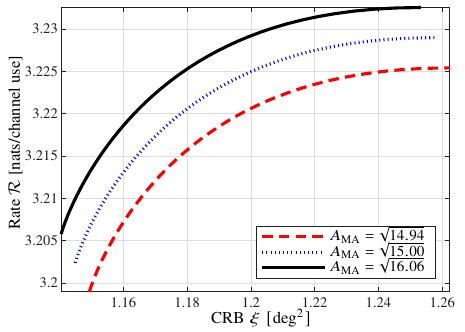}
	\caption{Pareto fronts for the CRB-Rate region for various $\Ama$.}
	\label{fig:Pareto_MA}
\end{figure}

For MA, the inner optimization tolerance and the outer KKT optimality tolerance are set to $\epsilon_{\mathrm{in}}=10^{-9}$ and $\epsilon_{\mathrm{out}}=2\times10^{-5}$, respectively, while the mass-point locations are refined using MATLAB's $\texttt{fmincon}()$ function with the sequential quadratic programming algorithm.

Consider $\Ama=\sqrt{15}$, we can calculate that $\alphas=0.576$. By changing $\alpha\in(\alphas,1]$, we can obtain the optimized $\Psopt$ in Fig. \ref{fig:PMF_MA}. As theoretically established in Theorem \ref{theorem:MA_Ps}, the maximum power $\s=A^2$ always belongs to the support of the optimal distribution $\Psopt$. As $\alpha$ increases, the probability mass assigned to $\s=A^2$ gradually decreases, while additional mass points emerge and receive increasing probability, reflecting the growing emphasis on communication preference. Consequently, the optimal distribution evolves from a single mass point at $\alpha\in[0,\alphas]$ to a richer discrete distribution as $\alpha$ approaches $1$.

The above observations are further corroborated by Fig. \ref{fig:mass_points_MA}, which illustrates the number of mass points $K_\alpha$ versus the sensing-communication preference $\alpha$ under different amplitude constraints $\Ama$. For a given $\Ama$, $K_\alpha$ increases in a staircase-like manner with $\alpha$, reflecting the gradual transition from a sensing-optimal single-mass-point distribution to a richer discrete distribution that supports higher communication rates. Comparing different amplitude constraints, we observe that a larger $\Ama$ generally leads to a larger number of mass points at $\alpha=1$, consistent with the classical information-theoretic characterization of amplitude-constrained Gaussian channels \cite{SMITH1971}. Moreover, the threshold $\alphas$, at which the optimal distribution transitions from one to multiple (at least two) mass points, increases with $\Ama$. This indicates that a larger amplitude limit allows the sensing-optimal single-mass-point distribution to remain optimal over a wider range of sensing-communication preferences, thereby delaying the emergence of additional mass points. Finally, we obtain the Pareto boundaries of the CRB-Rate regions under different amplitude constraints in Fig. \ref{fig:Pareto_MA}. As expected, increasing $A_{\mathrm{MA}}$ enlarges the achievable performance region, enabling higher communication rates for a given sensing CRB.

\section{Conclusion}
\label{sec:conclusion}

In this paper, we have investigated the fundamental estimation-information performance limits of ISAC systems, with particular emphasis on the structural properties of Pareto-optimal input distributions under different power constraints. By adopting the Miller-Chang-type CRB as the sensing metric, we have formulated the CRB-Rate Pareto optimization problem and shown that the optimal input distributions admit a common structure: the directional component is uniformly distributed over the complex unit sphere, while the power component exhibits distinct support characteristics depending on the power constraints and sensing-communication tradeoff preference. For GP and MA, we have studied their support characteristics in detail. We have further developed numerical algorithms to determine the optimal power distributions and trace the corresponding Pareto boundaries under GP and MA constraints, with numerical results corroborating our theoretical findings. Importantly, our results reveal that sensing favors deterministic transmit power, whereas communications benefit from increasingly diverse power levels, thereby providing a fundamental characterization of the waveform uncertainty tradeoff in the power domain and shedding light on the design of Pareto-optimal ISAC signaling strategies.

Several open problems remain for future investigation. These include the exact characterization of the transition between finite and infinite support under GP, whether the support cardinality under MA is monotonic with the sensing-communication preference, and the extension of the present structural analysis to general MIMO ISAC systems with spatially coupled waveform distributions.

\appendices
\numberwithin{equation}{section}

\section{Proof of Proposition \ref{proposition:closed_CRB}}
\label{appendix:closed_CRB}

\begin{proof}
We first derive the Bayesian \ac{FIM} according to the definition of the Bayesian \ac{CRB}. Given measurement model \eqref{eqn:measurement_model_s}:
\[
\ys:=\vect(\Ys)
=\hs(\boldeta)\otimes\x+\boldsymbol{\mathsf z}_{\rm s},
\]
where $\otimes$ is the Kronecker product, and $\boldsymbol{\mathsf z}_{\rm s}:=\vect(\Zs)\sim
\mathcal{CN}(\boldsymbol 0,\noisevar\I_{N\Lhs})$ is the equivalent noise. The Jacobian matrix of the channel response $\hs$ with respect to the (real-valued) parameter \(\boldeta\) is denoted as $\nabla_{\boldeta}\hs$ in the Proposition statement, which is inherent \emph{random} that is explicitly dependent on $\boldeta$. For any $\boldeta$ and a known waveform \(\x\), the Slepian-Bangs formula \cite{wang2025device,wang2026unifiedcramerraoboundframework,wang_holigraphic_TWC} yields the conditional FIM as
\begin{equation*}
	\begin{aligned}
		\Fr_{\mathrm{conditional}}(\boldeta|\x)
		&=
		\frac{2}{\noisevar}
		\Re\left\{
		\left[
		(\nabla_{\boldeta}\hs)
		\otimes\x
		\right]^\herm
		\left[
		(\nabla_{\boldeta}\hs)
		\otimes\x
		\right]
		\right\}\\
		&=
		\frac{2\|\x\|^2}{\noisevar}
		\Re\left\{
		(\nabla_{\boldeta}\hs)^\herm (\nabla_{\boldeta}\hs)
		\right\}.
	\end{aligned}
\end{equation*}
Since \(\boldeta\) is a random parameter, the Bayesian \ac{FIM} is obtained by averaging the conditional \ac{FIM} over the prior distribution and incorporating the prior information matrix \cite{wang_holigraphic_TWC}:
\begin{equation*}
	\begin{aligned}
		\Fr_{\mathrm{Bayesian}}(\x)
		&=
		\Fr(\x)
		+\Fp\\
		&=
		\mathbb E_{\boldeta}
		\left[
		\Fr_{\mathrm{conditional}}(\boldeta|\x)
		\right]
		+\Fp\\
		&=
		\|\x\|^2\tilde{\boldsymbol G}+\Fp,
	\end{aligned}
\end{equation*}
where
\begin{equation*}
	\tilde{\boldsymbol G}:= \frac{2}{\noisevar}
	\mathbb{E}\left[
	\Re\left\{
	(\nabla_{\boldeta}\hs)^\herm (\nabla_{\boldeta}\hs)\right\}\right],
\end{equation*}
Here, \(\tilde{\boldsymbol G}\) is a real symmetric positive semidefinite matrix. According to the definition of the Bayesian \ac{CRB}, the corresponding \ac{CRB} matrix for $\boldeta$ \eqref{eqn:CRB_def} is given by
\begin{equation*}
	\CRBm(\x)
	= \Fr_{\mathrm{Bayesian}}^{-1}(\x)=
	\left(
	\Fp+\|\x\|^2\tilde{\boldsymbol G}
	\right)^{-1}.
\end{equation*}

Next, define the whitened conditional FIM as
\[
\boldsymbol G:=
\Fp^{-1/2}
\tilde{\boldsymbol G}
\Fp^{-1/2},
\]
which is exactly \eqref{eqn:G_def}. Since \(\boldsymbol G\) is real symmetric positive semidefinite, it admits the eigendecomposition \eqref{eqn:G_decomp}. Hence, for any $\x$, the weighted sensing error can be expressed as
\[
\begin{aligned}
	&\ \tr\left\{
	\W
	\left(
	\Fp+\|\x\|^2\tilde{\boldsymbol G}
	\right)^{-1}
	\right\}\\
	=&\
	\tr\left\{
	\Fp^{-1/2}
	\W
	\Fp^{-1/2}
	\boldsymbol Q
	\left(
	\boldsymbol I+\|\x\|^2\boldSigma
	\right)^{-1}
	\boldsymbol Q^\trans
	\right\}\\
	=&\
	\tr\left\{
	\boldsymbol Q^\trans
	\Fp^{-1/2}
	\W
	\Fp^{-1/2}
	\boldsymbol Q
	\left(
	\boldsymbol I+\|\x\|^2\boldSigma
	\right)^{-1}
	\right\}\\
	=&\
	\tr\left\{
	\boldLambda
	\left(
	\boldsymbol I+\|\x\|^2\boldSigma
	\right)^{-1}
	\right\},
\end{aligned}
\]
where the last equality holds according to the definition \eqref{eqn:Lambda_def}. Notice that the matrix $(\I+\|\x\|^2\boldSigma)^{-1}$ is a diagonal matrix. From the property of the $\tr(\cdot)$ operator and taking the expectation over the distribution of $\x$, the weighted sensing error therefore admits closed-form expression \eqref{eqn:sensing_error} and \eqref{eqn:phi_def}.
\end{proof}
\section{Proof of Proposition \ref{proposition:KKT_GP}}
\label{appendix:KKT_GP}

\begin{proof}	
	Under the general power constraint \eqref{eqn:GP}, the optimization problem \eqref{eqn:Pareto} is given by
	\begin{equation*}
		\begin{aligned}
			\max_{\Ps}\quad&
			\Fobj[\Ps] =  \MI[\Ps]+\beta\xi[\Ps]\\
			\mathrm{s.t.}\quad&
			\int_0^\infty s\,\diff \Ps(s)\leq N\rho.
		\end{aligned}
	\end{equation*}
	Introducing the Lagrange multipliers $\nu_\alpha\geq0$ and $c_\alpha'\in\mathbb R$ for the average power and probability normalization constraints, respectively, the Lagrangian is
	\begin{equation}
		\begin{aligned}
			&\mathcal L_\alpha[\Ps]
			=
			\MI[\Ps]
			+\beta\int_0^\infty\phi(s)\,\diff \Ps(s)\\
			&-\nu_\alpha\left(
			\int_0^\infty s\,\diff \Ps(s)-N\rho
			\right) -c_\alpha'\left(
			\int_0^\infty \diff\Ps(s)-1
			\right).
		\end{aligned}
		\label{eqn:GP_Lagrangian}
	\end{equation}
	
	According to the definition of mutual information,
	\begin{equation*}
		\begin{aligned}
			\MI[\Ps] &= \MI(\x;\yc) = h(\yc)-h(\zc) \\
			&=
			-\int p_{\yc}(\yr;\Ps)
			\log p_{\yc}(\yr;\Ps)
			\,\diff\yr
			-h(\zc),
		\end{aligned}
	\end{equation*}
	where $h(\cdot)$ denotes the differential entropy, and
	\begin{equation*}
		p_{\yc}(\yr;\Ps)
		=
		\int_0^\infty
		p_{\yc|\s}(\yr|s)
		\,\diff \Ps(s).
	\end{equation*} 
	Taking the functional derivative with respect to
	$\Ps$ yields
	\begin{equation*}
		\begin{aligned}
			\frac{\delta}{\delta\Ps}\MI[\Ps]
			&=
			-\int p_{\yc|\s}(\yr|s)
			\left[
			\log p_{\yc}(\yr;\Ps)+1
			\right]\diff \yr\\
			&=
			\kappa(s;\Ps)-1.
		\end{aligned}
	\end{equation*}
	The first-order functional derivative of \eqref{eqn:GP_Lagrangian} is thus
	\begin{equation*}
		l_\alpha(s;\Ps) := \frac{\delta}{\delta\Ps}\mathcal L_\alpha[\Ps]
		=
		\kappa(s;\Ps)
		+\beta\phi(s)-\nu_\alpha s-c_\alpha,
		\label{eqn:GP_variation}
	\end{equation*}
	where $c_\alpha=1+c_\alpha'$. This yields \eqref{eqn:KKT_density}.
	
	Since the objective $\Fobj[\Ps]$ is concave in $\Ps$, the power constraint is affine and satisfies Slater's condition for $N\rho>0$, and an optimal solution exists by Lemma \ref{lemma:existance_Psopt}, the KKT conditions are necessary and sufficient. In particular, the first-order optimality condition yields
	\[
	l_\alpha(s;\Psopt)
	\le 0,\qquad \forall s\ge0,
	\]
	with equality $\Psopt$-almost surely. Conversely, concavity ensures that these conditions are sufficient for global optimality.
\end{proof}

\section{Useful Bounds on the (Conditional) Gaussian Output Probability Density Functions}

\begin{lemma}
	\label{lemma:Gaussian_output_PDF_bounds}
	
	Consider the equivalent Gaussian channel in \eqref{eqn:y_Yc}. For any power distribution $\Ps$ and any $M>0$ such that $\Ps([0,M])>0$, the output PDF satisfies
	\begin{equation}
		\begin{aligned}
			p_{\yc}(\yr;\Ps)
			&\ge \frac{\Ps([0,M])}{\pi^N}
			\exp\left[-\left(\|\yr\|+\sqrt{\SNRc M}\right)^2\right].
		\end{aligned}
		\label{eq:gaussian_pdf_lower_bound}
	\end{equation}
	Furthermore, if $\supp(\Ps)\subseteq[0,M]$, then
	\begin{equation}
		\begin{aligned}
			\frac{1}{\pi^N}
			\mathrm{e}^{-(\|\yr\|+\sqrt{\SNRc M})^2} \le p_{\yc}(\yr;\Ps) \le\frac{1}{\pi^N}
			\mathrm{e}^{-(\{\|\yr\|-\sqrt{\SNRc M}\}^+)^2}.
		\end{aligned}
		\label{eq:gaussian_pdf_two_sided_bound}
	\end{equation}
	
	\begin{proof}
		For any $s\in[0,M]$ and $\|\u^\star\|=1$, the triangle and reverse triangle inequalities yield
		\begin{equation}
			\left(\big\{\|\yr\|-\sqrt{\SNRc M}\big\}^+\right)^2
			\le \|\yr-\sqrt{\SNRc s}\u^\star \|
			\le \|\yr\|+\sqrt{\SNRc M}.
			\label{eq:gaussian_norm_two_sided_bound}
		\end{equation}
		Since the conditional Gaussian PDF is
		\begin{equation}
			p_{\yc|\s,\u}(\yr|s,\u^\star) = \pi^{-N}\exp(-\|\yr-\sqrt{\SNRc s}\u^\star\|^2),
			\label{eqn:PDF_y_condition_su}
		\end{equation}
		averaging the right inequality in \eqref{eq:gaussian_norm_two_sided_bound} over $s\in[0,M]$ and $\u^\star\sim\mu_N$ proves \eqref{eq:gaussian_pdf_lower_bound}. If $\supp(\Ps)\subseteq[0,M]$, averaging both bounds over the entire input distribution establishes \eqref{eq:gaussian_pdf_two_sided_bound}.
	\end{proof}
\end{lemma}

\section{Entire Analyticity of the Marginal\\ Entropy Density $\kappa(s;\Ps)$}
\label{appendix:kappa_analyticity}

In this section, we establish that the marginal entropy density $\kappa(s;\Ps)$ defined in Proposition \ref{proposition:KKT_GP} admits an entire analytic extension with respect to the power variable $s$, for any feasible $\Ps$. 

\begin{proof}	
	Given $\kappa(s;\Ps)$ in \eqref{eqn:kappa_def} defined as
	\begin{equation*}
		\kappa(s;\Ps)
		=
		-\int
		p_{\yc|\s}(\yr|s)
		\log p_{\yc}(\yr;\Ps)
		\,\diff\yr,
	\end{equation*}
	to show $\kappa(z;\Ps)$ is entire in $z\in \mathbb{C}$, we need to show that
	\begin{enumerate}
		\item[i)] for any fixed $\yr$, $z\mapsto p_{\yc|\s}(\yr|z)\log p_{\yc}(\yr;\Ps)$ is entire function;
		\item[ii)] for any compact set $U\subset\mathbb C$, the term
		\begin{equation*}
			\sup_{z\in U}\ |p_{\yc|\s}(\yr|z)
			\log p_{\yc}(\yr;\Ps)|
		\end{equation*}
		is bounded above by an integrable function of $\yr$ over $\mathbb C^N$.
	\end{enumerate}
	
	In what follows, we investigate the two terms $p_{\yc|\s}(\yr|z)$ and $\log p_{\yc}(\yr;\Ps)$, respectively.
	
	$\divideontimes$ The $p_{\yc|\s}(\yr|z)$-term. From \eqref{eqn:PDF_y_condition_su}, the conditional output PDF is
	\begin{equation*}
		\begin{aligned}
			p_{\yc|\s}(\yr|s)
			&=
			\frac{1}{\pi^N}
			\int_{\mathbb S_{\mathbb C}^{N-1}}
			\exp\left(
			-\left\|
			\yr-\sqrt{\SNRc s}\boldsymbol u
			\right\|^2
			\right)
			\,\diff\mu_N(\boldsymbol u)\\
			&=
			\frac{{\text e}^{-\|\yr\|^2-\SNRc s}}{\pi^N}
			\int_{\mathbb S_{\mathbb C}^{N-1}}
			\mathrm{e}^{2\sqrt{\SNRc s}
				\Re(\yr^\herm\boldsymbol u)}
			\,\diff\mu_N(\boldsymbol u).
		\end{aligned}
	\end{equation*}
	Using Taylor expansion of the exponential function, since the uniform spherical measure is invariant under $\boldsymbol u\mapsto-\boldsymbol u$, all odd-order moments vanish. Therefore,
	\begin{equation*}
		\begin{aligned}
			&p_{\yc|\s}(\yr|s)\\
			&=
			\frac{{\text e}^{-\|\yr\|^2-\SNRc s}}{\pi^N}
			\sum_{k=0}^{\infty}
			\frac{(4\SNRc s)^k}{(2k)!}
			\int_{\mathbb S_{\mathbb C}^{N-1}}
			\left[
			\Re(\yr^\herm\boldsymbol u)
			\right]^{2k}
			\,\diff\mu_N(\boldsymbol u).
		\end{aligned}
		\label{eqn:GP_power_series}
	\end{equation*}
	
	The right-hand side defines an analytic extension to $z\in\mathbb C$ by replacing $s$ with $z$.
	Indeed, using $\left| \Re(\yr^\herm\boldsymbol u) \right| \leq\|\yr\|$, and hence
	\begin{equation*}
		\int_{\mathbb S_{\mathbb C}^{N-1}}
		\left[
		\Re(\yr^\herm\boldsymbol u)
		\right]^{2k}
		\,\diff\mu_N(\boldsymbol u) 
		\leq 
		\int_{\mathbb S_{\mathbb C}^{N-1}}
		\|\yr\|^{2k}
		\,\diff\mu_N(\boldsymbol u) 
		= 
		\|\yr\|^{2k},
	\end{equation*}
	the absolute value of the series is up-bounded by
	\[
	\sum_{k=0}^{\infty}
	\frac{
		(4\SNRc|z|\|\yr\|^2)^k
	}{(2k)!}
	=
	\cosh\left(
	2\sqrt{\SNRc|z|}\|\yr\|
	\right).
	\]
	Thus, the series converges locally uniformly for every $z\in\mathbb C$, and $p_{\yc|\s}(\yr|z)$ is entire in $z$. Next, for any compact set $U\subset\mathbb C$,
	$$\begin{aligned}
		\cosh\!\left(
		2\sqrt{\SNRc|z|}\|\yr\|
		\right)
		&\leq
		\cosh\!\left(
		2\sqrt{\SNRc \sup_{z\in U} |z| }\|\yr\|
		\right)\\
		&\leq 
		\exp(2\sqrt{\SNRc  \sup_{z\in U}|z|}\|\yr\|),
	\end{aligned}$$
	and hence
	\begin{equation}
		\left|
		p_{\yc|\s}
		(\yr|z)
		\right|
		\leq
		C_U
		\exp\left[
		-\|\yr\|^2
		+
		B_U\|\yr\|
		\right],
		\label{eqn:GP_kernel_bound}
	\end{equation}
	where $C_U := \frac{{\text e}^{\SNRc \sup_{z\in U}|z|}}{\pi^N}$ and $B_U :=2\sqrt{\SNRc\sup_{z\in U}|z|}$. 
	
	$\divideontimes$ The $\log p_{\yc}(\yr;\Ps)$-term. From \eqref{eq:gaussian_pdf_lower_bound},
	\begin{equation*}
		\begin{aligned}
			-\log& p_{\yc}(\yr;\Ps)\\
			&\le
			-\log \Ps([0,M])+N\log\pi
			+\bigl(\|\yr\|+\sqrt{\SNRc M}\bigr)^2 \\
			&\le
			-\log \Ps([0,M])+N\log\pi
			+2\SNRc M+2\|\yr\|^2.
		\end{aligned}
	\end{equation*}

	Moreover,
	\begin{equation*}
		\begin{aligned}
			p_{\yc}(\yr;\Ps)
			&=
			\frac{1}{\pi^N}
			\iint_{\mathbb S_{\mathbb C}^{N-1}}
			\exp\left(
			-\left\|
			\yr-\sqrt{\SNRc s}\boldsymbol u
			\right\|^2
			\right)
			\,\diff\mu_N(\boldsymbol u)\diff\Ps\\
			&\leq 		\frac{1}{\pi^N}
			\iint_{\mathbb S_{\mathbb C}^{N-1}}
			1
			\,\diff\mu_N(\boldsymbol u) \diff \Ps = \pi^{-N}<1.
		\end{aligned}
	\end{equation*}
	Hence, $\log p_{\yc}(\yr;\Ps)<0$. We can conclude that there exists finite constant \(C_0= -\log \Ps([0,M]) +N\log\pi +2\SNRc M >0\), independent of \(\yr\), such that
	\begin{equation}
		\left|
		\log p_{\yc}
		(\yr;\Ps)
		\right| = -\log p_{\yc}
		(\yr;\Ps) \leq
		C_0+2\|\yr\|^2.
		\label{eqn:GP_log_bound}
	\end{equation}
	
	For condition i), as $p_{\yc|\s}(\yr|z)$ is entire and $\log p_{\yc}(\yr;\Ps)$ is independent of $z$, their product is entire. For condition ii), combining \eqref{eqn:GP_kernel_bound} and \eqref{eqn:GP_log_bound}, we obtain
	\begin{equation*}
		\begin{aligned}
			\sup_{z\in U}\quad 
			&\left|
			p_{\yc|\s}
			(\yr|z)
			\log p_{\yc}
			(\yr;\Ps)
			\right|\\
			&\leq
			C_U(C_0+2\|\yr\|^2)
			{\text e}^{-\|\yr\|^2+B_U\|\yr\|}.
		\end{aligned}
	\end{equation*}
	The right-hand side is clearly integrable over $\mathbb C^N$. Therefore, by the holomorphic parameter integral theorem (see, e.g., \cite[Chapter 2]{Stein2003ComplexAnalysis}),
	\[ 
	\kappa(z;\Ps) = -\int_{\mathbb C^N}
	p_{\yc|\s}(\yr|z)
	\log p_{\yc}
	(\yr;\Ps)
	\,\diff\yr
	\]
	is entire in $z$.
\end{proof}

\section{Proof of Theorem \ref{theorem:discreteness_GP}}
\label{appendix:discreteness_GP}

\begin{proof}	
	We first establish the following two facts
	\begin{enumerate}
		\item As established Appendix \ref{appendix:kappa_analyticity}, the marginal entropy density $\kappa(s;\Psopt)$ defined in \eqref{eqn:kappa_def} admits an entire analytic extension $\kappa(z;\Psopt)$ to $z\in\mathbb C$.
		\item  The sensing distortion
		\begin{equation*}
			\phi(z)
			=
			\sum_{\ell=1}^{\Ln}
			\frac{\lambda_\ell}{1+\sigma_\ell z}
		\end{equation*}
		has poles at $z=-1/\sigma_\ell<0$ when $\lambda_\ell\sigma_\ell>0$. 
		According to \eqref{eqn:sensing_condition}, at least one such pole exists and is non-removable. Hence, $\phi(z)$ is meromorphic but not entire.
	\end{enumerate}
	
	We prove the theorem by contradiction. Suppose that the optimal support has a finite accumulation point. According to Proposition \ref{proposition:KKT_GP}, the optimal power distribution satisfies
	\begin{equation*}
		l_\alpha(s;\Psopt)=0, \qquad  \forall s\in\supp(\Psopt).
		\label{eqn:GP_support_equality}
	\end{equation*}
	Since any finite accumulation point belongs to $[0,\infty)$, 
	in which both $\kappa(z;\Psopt)$ and $\phi(z)$ are holomorphic, the identity theorem \cite{Knopp1945} implies
	\begin{equation*}
		\kappa(z;\Psopt)
		-\nu_\alpha z-c_\alpha
		=
		-\beta\phi(z)
		\label{eqn:GP_analytic_identity}
	\end{equation*}
	throughout $\mathbb{C}$ excluding the poles of $\phi(z)$. However, as $z$ approaches any non-removable pole of \(\phi(z)\), the left-hand side remains finite (as it is entire) whereas the right-hand side diverges for $\beta\neq0$ (or equivalently, $0< \alpha< 1$), yielding a contradiction. Therefore, the optimal support has no finite accumulation points.\footnote{Note that accumulation at infinity does not satisfy the hypothesis of the identity theorem, as infinity lies outside the complex plane.} 
	
	Since the support is closed and has no finite accumulation points, by the Bolzano-Weierstrass theorem, its intersection with any compact interval is finite. Consequently, the support is at most countable, and the optimal power distribution is purely atomic, i.e., a finite or countable sum of Dirac measures, yielding \eqref{eqn:GP_discrete} and \eqref{eqn:GP_local_finite}.
	
	For $\alpha=0$, from Lemma \ref{lemma:corner_points_GP} we know that $\mathbb{P}^\star_{\s,0}=\delta_B$ and thus $K_0=1$. It is clear that \eqref{eqn:GP_discrete} and \eqref{eqn:GP_local_finite} also hold. 
\end{proof}

\section{Proof of Theorem \ref{theorem:K_alpha_GP}}
\label{appendix:K_alpha_GP}

	\subsection{Proof of \eqref{eqn:alpha_s} }
	
	\begin{proof}
	We first establish the single-mass-point optimality threshold $\alphas$. The aim is to find the preference threshold $\alphas\in(0,1)$ such that candidate distribution $\Psopt=\delta_B$, where $B=N\rho$, always holds for $\alpha\in[0,\alphas]$. As $B>0$, the KKT equality and stationarity at $s=B$ in \eqref{eqn:KKT_GP} yield
	\begin{equation*}
		\begin{aligned}
			\nu_\alpha &= \kappa'(B;\delta_B)+\beta\phi'(B);\\
			c_\alpha &= \kappa(B;\delta_B) + \beta\phi(B)-\nu_\alpha B.
		\end{aligned}
		\label{eqn:GP_nu_single}
	\end{equation*}
	Substituting them back into \eqref{eqn:KKT_GP} gives
	\begin{equation}
		D_\kappa(s)+\beta D_\phi(s)\leq0,
		\qquad \forall s\geq0.
		\label{eqn:GP_single_KKT}
	\end{equation}
	Since $\phi(s)$ is strictly convex, $D_\phi(s)>0$ for every $s\neq B$. Consequently, \eqref{eqn:GP_single_KKT} is equivalent to
	
	\begin{equation}
		\begin{aligned}
			&-\beta\geq \frac{D_\kappa(s)}{D_\phi(s)},
			\quad \forall s\geq0,\ s\neq B\\
			&\Longleftrightarrow-\beta\geq \sup_{\substack{s\geq0\\s\neq B}}
			\frac{D_\kappa(s)}{D_\phi(s)}=T(B).
		\end{aligned}
		\label{eqn:GP_single_condition}
	\end{equation}
	
	Recall $\beta\in(-\infty,0]$, we must have $0<T(B)<\infty$ to make sure the above inequality is meaningful. Since $D_\phi(s)>0$ for all $s\neq B$, the ratio
	$D_\kappa(s)/D_\phi(s)$ is continuous at every finite $s\neq B$. Thus, it suffices to examine its behavior as $s\to B$ and $s\to\infty$.
	
	For $s\to B$, by analyticity,
	\begin{equation*}
		\lim_{s\to B}
		\frac{D_\kappa(s)}{D_\phi(s)}
		=
		\frac{\kappa''(B;\delta_B)}{\phi''(B)},
	\end{equation*}
	which is finite as $\kappa(\cdot;\delta_B)$ is entire analytic and $0<\phi''(B)<\infty$. 
	
	For $s\to\infty$, we next characterize the asymptotic behavior of $\displaystyle\lim_{s\to\infty} \frac{D_\kappa(s)}{D_\phi (s)}$. First we consider the behavior of
	\begin{equation}
		\kappa(s;\delta_B) = 
		\mathbb E\left[
		-\log p_{\yc}(\yc;\delta_B)
		\mid \s = s
		\right],
		\label{eqn:kappa_expectation}
	\end{equation}
	as $s\to\infty$.
	
Recall the equivalent communication channel \eqref{eqn:y_Yc} is 
$$\yc=\sqrt{\SNRc \s}\,\u^\star+\zc,$$
where $\u^\star\sim\mu_N$. For $\Ps=\delta_B$, the output \ac{PDF} is
\begin{equation}
	\begin{aligned}
		& \ p_{\yc}(\yr;\delta_B)
		=p_{\yc|\s}(\yr|B)\\
		=&\ \frac{1}{\pi^N}
		\int_{\mathbb S_{\mathbb C}^{N-1}}
		\mathrm{e}^{-\|\yr-\sqrt{\SNRc B}\boldsymbol{u}\|^2}
		\,\diff\mu_N(\boldsymbol{u})\\
		=&\ \frac{\mathrm{e}^{-\|\yr\|^2-\SNRc B}}{\pi^N}
		\int_{\mathbb S_{\mathbb C}^{N-1}}
		\mathrm{e}^{2\sqrt{\SNRc B}\Re(\yr^\herm\boldsymbol{u})}
		\,\diff \mu_N(\boldsymbol{u})\\
		=&\ \frac{\Gamma(N)}{\pi^N} \mathrm{e}^{-\|\yr\|^2-\SNRc B}\left(\sqrt{\SNRc B}\|\yr\|\right)^{1-N}\!\! I_{N-1}\left(2\sqrt{\SNRc B}\|\yr\|\right),
	\end{aligned}
	\label{eqn:PDF_y_con_B}
\end{equation}
where $\Gamma(\cdot)$ is the Gamma function and $I_{N-1}(\cdot)$ is the modified Bessel function of the first kind. The last equality follows from the rotational invariance of the uniform spherical measure and the integral representation of the modified Bessel function \cite[Eq. (10.32.2)]{NIST_DLMF}. Together with the large-argument asymptotic expansion of the modified Bessel function
\cite[Eq.~(10.40.1)]{NIST_DLMF},
\begin{equation*}
	I_{N-1}(z)
	=
	\frac{\mathrm{e}^z}{\sqrt{2\pi z}}
	\left(1+O(z^{-1})\right),
	\qquad z\to\infty,
\end{equation*}
we obtain
\begin{equation}
	\begin{aligned}
		-\log p_{\yc}(\yr;\delta_B)
		={}&\|\yr\|^2
		-2\sqrt{\SNRc B}\|\yr\|\\
		&+\left(N-\frac12\right)
		\log\|\yr\|+O(1),
	\end{aligned}
	\label{eqn:GP_shell_log_asymptotic}
\end{equation}
as $\|\yr\|\to\infty$. Thus, we consider the decomposition
\begin{equation}
	-\log p_{\yc}(\yr;\delta_B) = 
	\|\yr\|^2-2\sqrt{\SNRc B}\,\|\yr\| + R(\|\yr\|),
	\label{eqn:logpy_decomp}
\end{equation}
where $R(\|\yr\|):=-\log p_{\yc}(\yr;\delta_B)-\|\yr\|^2+2\sqrt{\SNRc B}\,\|\yr\|$. To calculate \eqref{eqn:kappa_expectation}, from \eqref{eqn:logpy_decomp}, we need to compute the conditional expectation of terms $\|\yr\|^2$, $\|\yr\|$, and $R(\|\yr\|)$.

For a test power $s$, the conditional output satisfies
\begin{equation}
	\mathbb E[\|\yc\|^2\mid\s=s]
	=
	\SNRc s+N.
	\label{eqn:conditional_expectation_yc_square}
\end{equation}
Moreover, the reverse triangle inequality gives
\begin{equation*}
	\left|\|\yc\mid \s=s\|-\sqrt{\SNRc s}\right| = \left|
	\|\sqrt{\SNRc s}\,\u^\star+\zc\|
	-\sqrt{\SNRc s}
	\right|
	\leq\|\zc\|,
\end{equation*}
and hence
\begin{equation}
	\mathbb E[\|\yc\|\mid\s=s]
	=
	\sqrt{\SNRc s}+O(1).
	\label{eqn:conditional_expectation_yc}
\end{equation}

For the $R(\|\yr\|)$ term, the asymptotic expansion \eqref{eqn:GP_shell_log_asymptotic} gives
\begin{equation*}
	R(\|\yr\|)=\left(N-\frac12\right)\log \|\yr\|+O(1),\quad \|\yr\|\to\infty.
\end{equation*}
Hence, there exist constants $r_0>0$ and $C_1>0$ such that $|R(\|\yr\|)|\le C_1\log(2+\|\yr\|)$ for all $\|\yr\| \ge r_0$. On the other hand, since $p_{\yc}(\yr;\delta_B)$ is strictly positive, continuous, and radially symmetric, $R(\|\yr\|)$ is continuous and thus bounded on the compact interval $[0,r_0]$. Noting that $\log(2+\|\yr\|)\ge\log 2>0$, we can enlarge $C_1$, if necessary, to obtain a finite constant $C>0$ satisfying
\begin{equation*}
	|R(\|\yr\|)|\leq C\log(2+\|\yr\|),
	\qquad \forall \|\yr\|\geq0.
	\label{eqn:GP_remainder_bound}
\end{equation*}

Hence, by Jensen's inequality,
\begin{equation}
	\begin{aligned}
		\mathbb{E}\left[ R(\|\yc\|) \mid \s=s \right] &\leq  \mathbb{E}\left[ |R(\|\yc\|)| \mid \s=s \right]  \\
		&\leq C \mathbb E[\log(2+\|\yc\|)\mid\s=s] \\
		& \leq
		C \log\left(2+\mathbb E[\|\yc\|\mid\s=s]\right)\\
		& \overset{\mathrm{\eqref{eqn:conditional_expectation_yc}}}{=} O(\log s).
	\end{aligned}
	\label{eqn:conditional_expectation_R}
\end{equation}
Consequently, by the definition of the marginal entropy density in \eqref{eqn:kappa_expectation}, and combining \eqref{eqn:conditional_expectation_yc_square}, \eqref{eqn:conditional_expectation_yc}, and \eqref{eqn:conditional_expectation_R}, we have
\begin{equation*}
	\begin{aligned}
		\kappa(s;\delta_B)
		&=\mathbb E[\|\yc\|^2\mid\s=s]
		-2\sqrt{\SNRc B}\,
		\mathbb E[\|\yc\|\mid\s=s]\\
		&\quad+\mathbb E[R(\|\yc\|)\mid\s=s]\\
		&=\SNRc s+N
		-2\sqrt{\SNRc B}
		\left(\sqrt{\SNRc s}+O(1)\right)
		+O(\log s)\\
		&=\SNRc s-2\SNRc\sqrt{B}\sqrt{s}
		+O(\log s),
	\end{aligned}
\end{equation*}
as $s\to\infty$. In particular, $\lim_{s\to\infty} {\kappa(s;\delta_B)}/{s}=\SNRc$. Since $\phi(s)$ is bounded, it follows from \eqref{eqn:Dk_Dphi_def} that
	\begin{equation}
		\lim_{s\to\infty}
		\frac{D_\kappa(s)}{D_\phi(s)}
		=
		\lim_{s\to\infty}
		\frac{D_\kappa(s)/s}{D_\phi(s)/s}
		=
		\frac{\SNRc-\kappa'(B;\delta_B)}
		{-\phi'(B)}.
		\label{eqn:D_ratio_lim}
	\end{equation}
We next investigate the property of $\kappa'(B;\delta_B)$. Consider
\[
	\MI(\x;\yc|\s=s)
	=-\int p_{\yc|\s}(\yr|s)
	\log p_{\yc|\s}(\yr|s)\,\diff \yr-h(\zc).
\]
Hence,
\[
\begin{aligned}
	\frac{\partial }{\partial s}\MI(\x;\yc|\s=s)
	={}&-\int
	\frac{\partial p_{\yc|\s}(\yr|s)}{\partial s}
	\log p_{\yc|\s}(\yr|s)\,\diff \yr\\
	&-\int
	\frac{\partial p_{\yc|\s}(\yr|s)}{\partial s}
	\,\diff \yr\\
	={}&-\int
	\frac{\partial p_{\yc|\s}(\yr|s)}{\partial s}
	\log p_{\yc|\s}(\yr|s)\,\diff\yr,
\end{aligned}
\]
Therefore,
\begin{equation*}
	\begin{aligned}
		\left.\frac{\partial \MI(\x;\yc|\s=s)}{\partial s}\right|_{s=B} &= -\int
		\left.\frac{\partial p_{\yc|\s}(\yr|s)}{\partial s}\right|_{s=B}
		\log p_{\yc}(\yr;\delta_B)\,\diff\yr \\
		&= \kappa'(B;\delta_B).
	\end{aligned}
\end{equation*}

From the celebrated I-\ac{MMSE} identity \cite{I-MMSE}
\begin{equation}
	\begin{aligned}
			\kappa'(B;\delta_B) &= 	\left.\frac{\partial \MI(\x;\yc|\s=s)}{\partial s}\right|_{s=B}\\
			 &= \SNRc\mmse(\u^\star|\yc,\s=B) \\
			&< \SNRc \lmmse(\u^\star|\yc,\s=B)\\
			&= \frac{\SNRc}{1+\rho\SNRc}< \SNRc.
	\end{aligned}
	\label{eqn:kappa_derivative_relax}
\end{equation}
The \ac{MMSE} is strictly smaller than the \ac{LMMSE}, as the two estimators are not identical. Specifically, the conditional mean estimator of the constant-modulus input $\u^\star$ is bounded, whereas its \ac{LMMSE} estimator, given as 
$$\hat{\u}^\star|[\yc,(\s=B)]=\frac{\sqrt{\SNRc B}}{N+\SNRc B}\yc,$$ 
is clearly unbounded under Gaussian noise.

Finally, we have $0\leq \kappa'(B;\delta_B)<\SNRc$, and hence 
		\begin{equation*}
		\lim_{s\to\infty}
		\frac{D_\kappa(s)}{D_\phi(s)}
		=
		\frac{\SNRc-\kappa'(B;\delta_B)}
		{-\phi'(B)}>0.
	\end{equation*}
Therefore, we have $0<T(B)<\infty$. Furthermore, $\kappa'(B;\delta_B)\geq0$, $\beta<0$, and $\phi'(B)<0$ imply $\nu_\alpha>0$. Hence, \eqref{eqn:GP_single_condition} is necessary and sufficient for $\delta_B$ to satisfy the KKT conditions. By the definition of $\beta(\alpha)$ in \eqref{eqn:beta_def}, this is equivalent to $0< \alpha\leq\alphas$. The strict concavity of mutual information ensures uniqueness. For $\alpha=0$, we know from Lemma \ref{lemma:corner_points_GP} that $\Psopt=\delta_B$. Therefore, $\Psopt=\delta_B$ if and only if $\alpha\in[0,\alphas]$.

Moreover, since every optimal distribution fully utilizes the average power budget, any single-mass-point optimal distribution (i.e., $K_\alpha=1$) must be $\Psopt=\delta_B$. Conversely, when $\Psopt=\delta_B$, we know that $K_\alpha=1$. Hence, $\Psopt=\delta_B$ if and only if $K_\alpha=1$. 

This proves \eqref{eqn:alpha_s}.
\end{proof}
	
\subsection{Proof of \eqref{eqn:alpha_c}}
\begin{proof}
Next,  we derive the infinitely-many-mass-point optimality threshold $\alphac$. Suppose, for contradiction, that the optimal power distribution has bounded support contained in $[0,M]$ for some finite $M>0$. 

Taking the negative logarithm and using $\big(\{x-a\}^+\big)^2\geq x^2-2ax$ for $x,a\geq0$ on the right inequality of \eqref{eq:gaussian_pdf_two_sided_bound}, we obtain
	\begin{equation*}
			-\log p_{\yc}(\yr;\Psopt)
			\geq
			N\log\pi+\|\yr\|^2
			-2\sqrt{\SNRc M}\|\yr\|.
	\end{equation*}
	Taking the conditional expectation for a test power $s$ gives
	\begin{equation*}
			\kappa(s;P_{\s,\alpha}^\star)
			\geq
			N\log\pi
			+\mathbb E[\|\yc\|^2\,|\,\s=s]
			-2\sqrt{\SNRc M}
			\mathbb E[\|\yc\|\,|\,\s=s].
	\end{equation*}
	Leveraging Jensen's inequality yields
	\begin{equation}
		\mathbb E[\|\yc\|\mid\s=s] \leq \sqrt{\mathbb E[\|\yc\|^2\mid\s=s]} \overset{\mathrm{\eqref{eqn:conditional_expectation_yc_square}}}{=} \sqrt{\SNRc s+N}.
		\label{eqn:conditional_expectation_yc_Jensen}
	\end{equation}
	Therefore,
	\begin{equation*}
		\kappa(s;P_{\s,\alpha}^\star)
			\geq
			\SNRc s+N+N\log\pi
			-2\sqrt{\SNRc M}\sqrt{\SNRc s+N}.
	\end{equation*}
	Dividing both sides by $s$ and taking $s\to\infty$, we obtain
	\begin{equation*}
		\liminf_{s\to\infty}
		\frac{\kappa(s;P_{\s,\alpha}^\star)}{s}
		\geq\SNRc.
		\label{eqn:GP_kappa_lower}
	\end{equation*}
	Moreover, the KKT inequality  in \eqref{eqn:KKT_GP} implies
	\begin{equation*}
		\nu_\alpha
		\geq
		\frac{\kappa(s;P_{\s,\alpha}^\star)}{s}
		+\frac{\beta\phi(s)-c_\alpha}{s},\qquad \forall s\geq 0.
	\end{equation*}
	As above inequality holds for every positive $s$, a necessary condition is 
	\begin{equation*}
		\nu_\alpha \geq \lim_{s\to\infty} \left[\frac{\kappa(s;P_{\s,\alpha}^\star)}{s}
		+\frac{\beta\phi(s)}{s}-\frac{c_\alpha}{s}\right] \geq \SNRc,
 	\end{equation*} 
	where the second inequality holds as $\phi(s)$ is bounded and $c_\alpha$ is finite. By now, we have shown that 
	\begin{equation}
		\supp(\Psopt) \subseteq [0,M] \Longrightarrow \nu_\alpha\geq \SNRc.
		\label{eqn:GP_nu_lower}
	\end{equation}
	
	On the other hand, the optimal distribution must fully utilize the average power budget, i.e., $\mathbb E[\s]=B=N\rho$. This is intuitive, as the mutual information and negative sensing distortion are non-decreasing with respect to the power. Consider the scaling perturbation $\x_a=\sqrt{a}\x$, where $a>0$. Along this perturbation, the Lagrangian, up to terms independent of $a$, is
	\begin{equation*}
		\mathcal L(a)
			={}\MI(\x;\sqrt{a\SNRc}\x+\zc)+\beta\mathbb E[\phi(a\s)]
			-\nu_\alpha(aB-B)+ \mathrm{const}.
	\end{equation*}
	At the optimal distribution, the first-order stationarity condition gives
	\begin{equation}
		\left.\frac{\partial \mathcal L(a)}{\partial a}
		\right|_{a=1}=0.
		\label{eqn:La_derivative_zero}
	\end{equation}
	By the I-\ac{MMSE} identity \cite{I-MMSE},
	\begin{equation}
		\left.
		\frac{\partial}{\partial a}
		\MI(\x;\sqrt{a\SNRc}\x+\zc)
		\right|_{a=1}
		=
		\SNRc\mmse(\x|\yc).
		\label{eqn:I_power_scale_derivative}
	\end{equation}
	Finally, \eqref{eqn:La_derivative_zero} gives
	\begin{equation*}
		\begin{aligned}
			0 &= 
			\SNRc\mmse(\x|\yc)
			+\beta\left.
			\frac{\partial }{\partial a}\mathbb E[\phi(a\s)]
			\right|_{a=1}
			-\nu_\alpha B \\
			&= \SNRc\mmse(\x|\yc)
			+\beta\mathbb E[\s\phi'(\s)]
			-\nu_\alpha B.
		\end{aligned}
	\end{equation*}

	Since the optimal input is isotropic, the \ac{LMMSE} bound gives
	\begin{equation*}
		\mmse(\x|\yc)
		\leq \lmmse(\x|\yc) = 
		\frac{N\rho}{1+\SNRc\rho}.
	\end{equation*}
	Therefore, for all feasible $\Psopt$,
	\begin{equation}
		\nu_\alpha
		\leq
		\frac{\SNRc}{1+\SNRc\rho}
		-\frac{\beta}{N\rho}\left(\sup_{s\geq 0}\{-s\phi'(s)\}\right).
		\label{eqn:GP_nu_upper}
	\end{equation}
	
	When $0<-\beta<D_{\mathrm{c}}$, the right-hand side of \eqref{eqn:GP_nu_upper} is strictly smaller than $\SNRc$, i.e.,
	\begin{equation*}
		0<-\beta<D_{\mathrm{c}} \Longleftrightarrow \alphac<\alpha<1 \Longrightarrow \nu_\alpha<\SNRc,
	\end{equation*}
	which contradicts \eqref{eqn:GP_nu_lower}. Thus, when $\alpha\in(\alphac,1)$ the optimal support must be unbounded. By Theorem \ref{theorem:discreteness_GP}, the optimal distribution must have countably infinitely many mass points whenever $\alpha\in(\alphac,1)$, proving \eqref{eqn:alpha_c}.

Finally, we establish the strict ordering of the two thresholds. 
\begin{equation*}
	\begin{aligned}
		T(B) = \sup_{\substack{s\geq0\\s\neq B}}
		\frac{D_\kappa(s)}{D_\phi(s)} 
		&\geq \lim_{s\to\infty} \frac{D_\kappa(s)}{D_\phi(s)}  \\
		& \overset{\mathrm{\eqref{eqn:D_ratio_lim}}}{=}
		 \frac{\SNRc-\kappa'(B;\delta_B)}
		{-\phi'(B)}\\
		& \overset{\mathrm{\eqref{eqn:kappa_derivative_relax}}}{>} \frac{(\SNRc-\frac{\SNRc}{1+\SNRc\rho})B}{-B\phi'(B)}\\
		&\geq 
		\frac{N\SNRc^2\rho^2}
		{(1+\SNRc\rho)
			\sup_{s\geq0}\{-s\phi'(s)\}}\\
		&=D_\mathrm{c}.
	\end{aligned}
\end{equation*}
Hence, $0<\alphas<\alphac<1$, which completes the proof. 
\end{proof}

\section{Proof of Corollary~\ref{cor:GP_alphac}}
\label{appendix:GP_alpha_c}
\begin{proof}
	Recall from Theorem \ref{theorem:K_alpha_GP} that
	\begin{equation*}
		\alphac
		=
		\left[
		1+
		\frac{\SNRc^2\rho^2}
		{(1+\SNRc\rho)\log(1+\SNRc\rho)}
		\frac{\phi(N\rho)}{A_\phi}
		\right]^{-1},
		\label{eqn:alphac_reformulated}
	\end{equation*}
	where
	\begin{equation*}
		A_\phi
		:= \sum_{s\geq 0}\ -s\phi'(s) = 
		\sup_{s\geq0}\ 
		\sum_{\ell=1}^{\Ln}
		\frac{\lambda_\ell\sigma_\ell s}
		{(1+\sigma_\ell s)^2}>0.
	\end{equation*}
	
	1) \emph{Communication channel gain:} 
	Let $t:=\SNRc\rho>0$ and define
	\begin{equation*}
		f(t):=\frac{t^2}{(1+t)\log(1+t)}>0,
	\end{equation*}
	whose logarithmic derivative is
	\begin{equation*}
		\frac{f'(t)}{f(t)}
		=
		\frac{(2+t)\log(1+t)-t}
		{t(1+t)\log(1+t)}.
	\end{equation*}
	Since $\log(1+t)>t/(1+t)$ for $t>0$, we have $f'(t)>0$. As $\phi(N\rho)/A_\phi$ is positive and independent of $\SNRc$, $\alphac$ is strictly decreasing in $\SNRc$.
	
	2) \emph{Input dimension:}
	For fixed $\rho$, the dependence on $N$ is through
	\begin{equation*}
		\phi(N\rho)
		=
		\sum_{\ell=1}^{\Ln}
		\frac{\lambda_\ell}{1+\sigma_\ell N\rho}.
	\end{equation*}
	This quantity is decreasing in $N$. Since $A_\phi$ is independent of $N$, the claimed monotonicity follows.
	
	3) \emph{Sensing distortion coefficients:}
	For any $a>0$, consider the common scaling $\lambda_\ell\mapsto a\lambda_\ell$ for all $\ell$ gives
	\begin{equation*}
		\phi(N\rho)\mapsto a\phi(N\rho),
		\qquad
		A_\phi\mapsto aA_\phi.
	\end{equation*}
	Hence, their ratio remains unchanged, and so does $\alphac$.
	
	4) \emph{Sensing information gains:}
	For any $a>0$, consider the common scaling $\sigma_\ell\mapsto a\sigma_\ell$ for all $\ell$. Then
	\begin{equation*}
		\begin{aligned}
			A_\phi
			=
			\sup_{s\geq0}
			\sum_{\ell=1}^{\Ln}
			\frac{\lambda_\ell a\sigma_\ell s}
			{(1+a\sigma_\ell s)^2}
			=
			\sup_{s'\geq0}
			\sum_{\ell=1}^{\Ln}
			\frac{\lambda_\ell\sigma_\ell s'}
			{(1+\sigma_\ell s')^2}
			\mapsto A_\phi,
		\end{aligned}
	\end{equation*}
	where $s'=as$. On the other hand,
	\begin{equation*}
		\phi(N\rho)
		=
		\sum_{\ell=1}^{\Ln}
		\frac{\lambda_\ell}
		{1+a\sigma_\ell N\rho}
	\end{equation*}
	is strictly decreasing in $a$. Therefore, $\alphac$ is increasing under common scaling of the sensing information gains.
	
	5) \emph{Transmit power:}
	As $\rho\to0^+$,
	\begin{equation*}
		\log(1+\SNRc\rho)
		=
		\SNRc\rho+O(\rho^2),
	\end{equation*}
	and $\phi(N\rho)\to\phi(0)<\infty$. Consequently,
	\begin{equation*}
		\lim_{\rho\to0}\frac{\SNRc^2\rho^2\phi(N\rho)}
		{(1+\SNRc\rho)\log(1+\SNRc\rho)A_\phi} = 0,
	\end{equation*}
	which implies $\alphac\to1$ as $\rho\to0$.
	
	Next, suppose $\sigma_\ell>0$ whenever
	$\lambda_\ell>0$. As $\rho\to\infty$,
	\begin{equation*}
		\begin{aligned}
			\phi(N\rho)
			&=
			\sum_{\ell=1}^{\Ln} \frac{\lambda_\ell}{1+\sigma_\ell N\rho}\\
			&= 
			\sum_{\ell=1}^{\Ln} \lambda_\ell \left(\frac{1}{N\sigma_\ell \rho}+O(\rho^{-2})\right) \\		
			&= \left(\frac{1}{N}
			\sum_{\ell=1}^{\Ln}
			\frac{\lambda_\ell}{\sigma_\ell}\right) \frac{1}{\rho}
			+O(\rho^{-2}).
		\end{aligned}
	\end{equation*}
	Furthermore,
	\begin{equation*}
		\frac{\SNRc^2\rho^2}
		{(1+\SNRc\rho)\log(1+\SNRc\rho)}
		\sim
		\frac{\SNRc\rho}
		{\log(\SNRc\rho)}.
	\end{equation*}
	It immediately follows that when $\rho\to\infty$
	\begin{equation*}
		\frac{\SNRc^2\rho^2\phi(N\rho)}
		{(1+\SNRc\rho)\log(1+\SNRc\rho)A_\phi}
		\sim
		\frac{\SNRc}
		{NA_\phi\log(\SNRc\rho)}
		\sum_{\ell=1}^{\Ln}
		\frac{\lambda_\ell}{\sigma_\ell}
		\to0.
	\end{equation*}
	Thus, $\alphac\to1$ as $\rho\to\infty$.
	
	Finally, $\alphac(\rho)$ is continuous and satisfies $0<\alphac(\rho)<1$ for every finite $\rho>0$, while approaching one at both endpoints. Therefore, it attains at least one global minimum in $(0,\infty)$ and cannot be monotonic.
\end{proof}
\section{Proof of Theorem \ref{theorem:MA_discrete}}
\label{appendix:MA_discrete}

\begin{proof}
	The proof follows the same analytic-function argument as Theorem \ref{theorem:discreteness_GP}. Specifically, the KKT equality in Proposition~\ref{proposition:MA_KKT} cannot hold on a support set with a finite accumulation point, since the marginal entropy density $\kappa(\cdot)$ is entire analytic, whereas $\phi(\cdot)$ has at least one non-removable pole for $\alpha\in(0,1)$ ($\beta <0$ and is finite). Hence, the optimal support has no finite accumulation points.  Suppose, for contradiction, that the optimal support contains infinitely many distinct points. Since it is contained in
	the compact interval $[0,A^2]$ under MA, the Bolzano-Weierstrass theorem guarantees the existence of a finite accumulation point, contradicting the preceding result. Therefore, the optimal support must be finite, and the optimal power distribution admits the representation \eqref{eqn:MA_discrete}.
	
	For $\alpha=1$ ($\beta=0$), the problem \eqref{eqn:Pareto} reduces to amplitude-constrained capacity problem, whose structural property of the optimal input distribution has been proven in \cite{SMITH1971}. 
	
	For $\alpha=0$ ($\beta\to-\infty$), we have shown in lemma \ref{lemma:corner_points_MA} that $P_{\s,0}^\star=\delta_{A^2}$. 
\end{proof}

\section{Proof of Theorem \ref{theorem:MA_Ps}}
\label{appendix:MA_Ps}
\begin{proof}
	In what follows, we prove these three properties separately.
	
	1) Suppose, by contradiction, that $A^2\notin\supp(\Psopt)$. Since the optimal distribution has finite support, there exists $s_{\max}:=\max\{\supp(\Psopt)\}\in(0,A^2)$.
	
	Choose a constant $a>1$ satisfying $a s_{\max}\le A^2$, and let $\widetilde{\Ps}$ denote the scaled distribution $\widetilde{\s}=a\s^\star$, where $\s^\star\sim\Psopt$. Clearly, $\widetilde{\Ps}$ remains feasible under the MA constraint. Given the equivalent channel input \eqref{eqn:y_Yc}, the derivative of the mutual information with respect to $a$, by the Gaussian I--MMSE identity \cite{I-MMSE}, is strictly positive (see, e.g., \eqref{eqn:I_power_scale_derivative}). Hence, $\MI[\widetilde{\Ps}]>\MI[\Psopt]$. Moreover, since $\phi(s)$ is strictly decreasing,
	$\xi[\widetilde{\Ps}] =\mathbb E[\phi(a\s^\star)] <\mathbb E[\phi(\s^\star)] =\xi[\Psopt]$. Thus, $\widetilde{\Ps}$ simultaneously improves the communication rate and sensing distortion, contradicting the Pareto-optimality of $\Psopt$. By contradiction, the argument is proven.
	
	2) When there is only a single mass point, we know from property 1) that $A^2\in\supp(\Psopt)$ and hence $\Psopt=\delta_{A^2}$. From the KKT condition \eqref{eqn:MA_KKT}:
	\begin{equation*}
		c_\alpha
		=
		\kappa(A^2;\delta_{A^2})+\beta\phi(A^2) \geq \kappa(s;\delta_{A^2}) + \beta\phi(s).
	\end{equation*}
	Recall that $\phi(s)$ is decreasing with $s$. Equivalently,
	\begin{equation*}
		-\beta \geq  \frac{\kappa(s,\delta_{A^2})-\kappa(A^2,\delta_{A^2})}{\phi(s)-\phi(A^2)}, \qquad \forall s\in[0,A^2].
	\end{equation*}
	Hence, we have
	\begin{equation*}
		\begin{aligned}
			-\beta \geq &\sup_{s\in[0,A^2]}  \frac{\kappa(s,\delta_{A^2})-\kappa(A^2,\delta_{A^2})}{\phi(s)-\phi(A^2)} = T(A^2) \\
			&\quad \Longleftrightarrow \alpha\in[0,\alphas],
		\end{aligned}
	\end{equation*}
	where $\alphas$ is given in \eqref{eqn:alpha_s_MA}. Recall that $-\beta\geq 0$, therefore we must project $T(A^2)$ to $\{T(A^2)\}^+$, yielding \eqref{eqn:T_def_MA}.
	
	Furthermore, the strict concavity of the mutual information functional ensures the uniqueness of the optimal input distribution for $\alpha>0$. Consequently, \eqref{eqn:single_mass_region_MA} can be proven.

	3) For $\alpha=0$, the result follows immediately from $\Psopt=\delta_{A^2}$.
	
	For $\alpha\in(0,1)$, consider the KKT density $l_\alpha(s;\Psopt)$ defined in \eqref{eqn:KKT_density_MA}, where $c_\alpha-N\log(\pi\mathrm{e})=\Fobj[\Psopt]=\MI[\Psopt]+\beta\xi[\Psopt]$. The KKT conditions require
	\begin{equation*}
		l_\alpha(s;\Psopt)\le0,
		\qquad \forall s\in[0,A^2],
	\end{equation*}
	with equality at every support point.
	
	Suppose that $0\in\supp(\Psopt)$. Then $l_\alpha(0;\Psopt)=0$, and the right-sided derivative must satisfy 
	\begin{equation}
		l_\alpha'(0^+;\Psopt)\leq0,
		\label{eqn:alpha0_l'}
	\end{equation}
	where the derivative is 
	\begin{equation}
		\frac{\partial}{\partial s}l_\alpha(s;\Ps)=\kappa'(s;\Ps)+\beta \phi'(s).
		\label{eqn:l_derivative}
	\end{equation}
	We next establish a distribution-independent lower bound on $\kappa'(0;\Ps)$.

	Recall the equivalent Gaussian channel $\yc=\sqrt{\SNRc \s}\u^\star+\zc$ in \eqref{eqn:y_Yc}. Define
	\begin{equation}
		\mathsf{t}:=2\|\yc\|^2.
		\label{eqn:t_transform}
	\end{equation}
	Conditioned on $\s=s$ and $\u^\star$, the vector $\yc$ is complex Gaussian with mean $\sqrt{\SNRc s}\,\u^\star$ and identity covariance. Therefore, for $s\neq 0$, the conditional distribution of $\mathsf{t}$ is a Chi-squared distribution with $2N$ degrees-of-freedom and non-centrality parameter $2\SNRc s$, i.e.,
	\begin{equation*}
		\mathsf{t}|(\s=s,\u^\star)
		\sim\chi^2_{2N}(2\SNRc s).
	\end{equation*}
	Since the non-centrality parameter is independent of $\u^\star$, averaging over the uniform directional distribution according to $\u^\star\sim\mu_N$ does not change this conditional law. Thus, 
	\begin{equation}
		\mathsf{t}|(\s=s)\sim\chi^2_{2N}(2\SNRc s).
		\label{eqn:t_conditional_PDF}
	\end{equation}
	For $t\geq 0$, the scalar conditional PDF is given as\footnote{Alternatively, we may simply replace $\|\yr\|$ with $\sqrt{t/2}$ and $B$ with $s$ in \eqref{eqn:PDF_y_con_B} and apply the chain rule for variable transformation to obtain \eqref{eq:app_scalar_pdf}.}
	\begin{equation}
		\begin{aligned}
			p_{\mathsf{t}|\s}(t|s)
			&=
			f_{\chi^2_{2N}(2\SNRc s)}(t) \\
			&= \frac12\mathrm{e}^{-t/2-\SNRc s}
			\left(\frac{t}{2\SNRc s}\right)^{
				\frac{N-1}{2}}
			 I_{N-1}
			\left(\sqrt{2\SNRc st}\right).
		\end{aligned}
		\label{eq:app_scalar_pdf}
	\end{equation}
	Specifically at $s=0$, $\mathsf{t}$ follows the central Chi-squared distribution with $2N$ degrees of freedom:
	\begin{equation*}
		p_{\mathsf{t}|\s}(t|0) = f_{\chi^2_{2N}(0)}(t)
		=\frac{t^{N-1}\mathrm{e}^{-t/2}}{2^N\Gamma(N)},\quad t\geq 0.
		\label{eq:radial_central_pdf}
	\end{equation*}
	
	Using the identity \cite[Eq. (10.39.9)]{NIST_DLMF}
	\begin{equation*}
		I_{N-1}(x)=
	\frac{(x/2)^\nu}{\Gamma(N)}
	{}_0F_1\left(;N;\frac{x^2}{4}\right),
	\end{equation*}
	where ${}_0F_1(;N;x)$ is the generalized hypergeometric function, the conditional PDF can equivalently be written as
	\begin{equation}
		p_{\mathsf{t}|\s}(t|s)
		=\frac{t^{N-1}}{2^N\Gamma(N)}
		\mathrm{e}^{-t/2-\SNRc  s}
		{}_0F_1\left(;N;\frac{\SNRc  st}{2}\right).
		\label{eq:radial_hypergeometric_pdf}
	\end{equation}
	Unlike the Bessel representation, this expression is well-defined at $s=0$. By the polar-coordinate change of variables in $\mathbb C^N$, we have
	\begin{equation*}
		\begin{aligned}
			p_{\yc|\s}(\yr|s)
			&=
			\frac{2^N\Gamma(N)}{\pi^N t^{N-1}}
			p_{\mathsf{t}|\s}(t|s)\\
			&=\pi^{-N}\mathrm{e}^{-t/2-\SNRc  s} 
			{}_0F_1\left(;N;\frac{\SNRc  st}{2}\right).
		\end{aligned}
		\label{eq:vector_conditional_hypergeometric}
	\end{equation*}
	Finally, averaging over the input power distribution $\Ps$ yields
	\begin{equation*}
		\begin{aligned}
			p_{\yc}(\yr;\Ps) =\int_0^{A^2}
			p_{\yc|\s}(\yr|s)\,\diff\Ps(s) =\pi^{-N}\mathrm{e}^{-t/2}M(t;\Ps),
		\end{aligned}
		\label{eq:vector_output_M}
	\end{equation*}
	where
	\begin{equation}
		M(t;\Ps):=
		\int_0^{A^2}
		\mathrm{e}^{-\SNRc  s}
		{}_0F_1\left(;N;\frac{\SNRc  st}{2}\right)
		\diff\Ps(s).
		\label{eq:M_definition}
	\end{equation}
	
	Consequently, the marginal entropy density in \eqref{eqn:kappa_def} becomes
	\begin{equation*}
		\kappa(s;\Ps)
		=N\log\pi+
		\int_0^\infty p_{\mathsf{t}|\s}(t|s)
		\left[\frac{t}{2}-\log M(t;\Ps)\right]\diff t.
		\label{eq:kappa_M}
	\end{equation*}
	
	Differentiating the conditional PDF with respect to $s$ and evaluating at $s=0$ give
	\begin{equation*}
		\left.
		\left[\frac{\partial}{\partial s}p_{\mathsf{t}|\s}(t|s)]\right]
		\right|_{s=0}
		=
		\SNRc  p_{\mathsf{t}|\s}(t|0)
		\left(\frac{t}{2N}-1\right),
		\label{eq:pt_derivative_zero}
	\end{equation*}
	where we used
	${}_0F_1(;N;0)=1$ and
	\begin{equation*}
		\left.\left[
		\frac{\partial}{\partial s}
		{}_0F_1\left(;N;\frac{\SNRc  st}{2}\right)\right]
		\right|_{s=0}
		=\frac{\SNRc  t}{2N},
	\end{equation*}
	which follow directly from the series representation \cite[Eq. (16.2.1)]{NIST_DLMF}. Finally, we obtain
	\begin{equation*}
		\kappa'(0;\Ps)
		=\SNRc \int_0^\infty p_{\mathsf{t}|\s}(t|0)
		\left(\frac{t}{2N}-1\right)
		\left[\frac{t}{2}-\log M(t;\Ps)\right]\diff t.
		\label{eq:kappa_zero_before_ibp}
	\end{equation*}
	
	Since
	\begin{equation*}
		p_{\mathsf{t}|\s}(t|0)
		\left(\frac{t}{2N}-1\right)
		=
		\frac{\partial}{\partial t} \left[-2f_{\chi^2_{2N+2}(0)}(t)\right],
	\end{equation*}
	where
	\begin{equation*}
		f_{\chi^2_{(2N+2)}(0)}(t)
		=\frac{t^N \mathrm{e}^{-t/2}}{2^{N+1}\Gamma(N+1)}
	\end{equation*}
	is the \ac{PDF} of a central Chi-squared distribution with $(2N+2)$ degrees-of-freedom, integration by parts yields \eqref{eq:kappa_zero_exact}.
	\begin{figure*}[t]
			\begin{equation}
			\begin{aligned}
				\kappa'(0;\Ps)
				& = \SNRc\Bigg[ -2f_{\chi^2_{(2N+2)}(0)}(t)\Big(\frac t2-\log M(t;\Ps)\Big)\Bigg|_0^\infty + 2 \int_0^\infty f_{\chi^2_{(2N+2)}(0)}(t)\Big(\frac12-\frac{M'(t;\Ps)}{M(t;\Ps)}\Big)\diff t \Bigg] \\
				&= \SNRc\Bigg[ 0+\int f_{\chi^2_{(2N+2)}(0)}(t)\,\diff t -2\int f_{\chi^2_{(2N+2)}(0)}(t) \frac{M'(t;\Ps)}{M(t;\Ps)} \diff t \Bigg] \\
				&=\SNRc \left[
				1-2\int_0^\infty f_{\chi^2_{(2N+2)}(0)}(t)
				\frac{M'(t;\Ps)}{M(t;\Ps)}\diff t
				\right]
			\end{aligned}
			\label{eq:kappa_zero_exact}
		\end{equation}
		\rule{\linewidth}{1pt}
	\end{figure*}
	Next, differentiating \eqref{eq:M_definition} gives
	\begin{equation*}
		M'(t;\Ps)
		=\int_0^{A^2}
		\frac{\SNRc  s}{2N}\mathrm{e}^{-\SNRc  s}
		{}_0F_1\left(;N+1;\frac{\SNRc  st}{2}\right)
		\diff\Ps(s).
	\end{equation*}
	Since $ {}_0F_1(;N+1;x) \leq  {}_0F_1(;N;x)$ for $x\ge0$, and $0\le s\le A^2$, it follows that
	\begin{equation*}
		0\le\frac{M'(t;\Ps)}{M(t;\Ps)}
		\le\frac{\SNRc  A^2}{2N}.
	\end{equation*}
	We finally obtain
	\begin{equation}
		\kappa'(0;\Ps) \ge\SNRc \left(1-\frac{\SNRc  A^2}{N}\right).
	\end{equation}
	
	From \eqref{eqn:l_derivative}:
	\begin{equation*}
		\begin{aligned}
			\left.\left[\frac{\partial}{\partial s}l_\alpha(s;\Psopt)\right]\right|_{s=0} &= \kappa'(0;\Psopt)+\beta \phi'(0)\\
			&\geq \SNRc \left(1-\frac{\SNRc  A^2}{N}\right)-\beta\sum_{\ell=1}^{\Ln}\sigma_\ell\lambda_\ell. 
		\end{aligned}
	\end{equation*}
	When 
	\begin{equation*}
		-\beta > \frac{\SNRc\left\{\frac{\SNRc A^2}{N}-1\right\}^+}{\sum_{\ell=1}^{\Ln}\sigma_\ell\lambda_\ell} \Longleftrightarrow 0\leq \alpha < \alpha_0,
	\end{equation*}
	we have $l'_{\alpha}(0;\Psopt)>0$, contradicting \eqref{eqn:alpha0_l'}. By contradiction, when $0\leq \alpha < \alpha_0$, we must have $0\notin\supp(\Psopt)$.
\end{proof}

\section{Proof of Theorem~\ref{thm:MA_cardinality_bound}}
\label{appendix:MA_cardinality_bound}

\begin{proof}
For $\alpha\in[0,\alphas]$, we know from Theorem \ref{theorem:MA_Ps} that a single mass point is optimal, i.e., $\Psopt=\delta_{A^2}$.

When $\alpha\in(\alphas,1)$, we know that $K_\alpha$ is at least 2. Hence, $2\leq \Kl\leq K\leq \Ku$. In what follows, we derive $\Kl$ and $\Ku$ for such $\alpha$.

\subsection{Lower Bound}

Note that $P_{\s,1}^\star$ is the communication-optimal power distribution under MA, i.e., the capacity-achieving distribution for \eqref{eqn:capacity_MA}. Hence,
\begin{equation*}
	\MI[P_{\s,1}^\star]=N\Rate^\star.
\end{equation*}
By the optimality of $\Psopt$, $\MI[\Psopt] +\beta\xi[\Psopt] \geq N\Rate^\star+\beta\xi[P_{\s,1}^\star]$, $\beta\leq 0$, and $\xi^\star\leq \xi[\Psopt]$, we have
\begin{equation}
	\MI[\Psopt]
	\geq N\Rate^\star+\beta(\xi[P_{\s,1}^\star]-\xi^\star).
	\label{eqn:MA_LB_1}
\end{equation}

Recall that $\x=\sqrt{\s}\u^\star$, where $\u^\star\sim\mu_N$ is independent of $\s$. By the chain rule,
\begin{equation}
	\begin{aligned}
		\MI[\Psopt]
		&=\MI(\s;\yc)+\MI(\u^\star;\yc|\s)\\
		&\leq \mathcal{H}(\s)+\MI[\delta_{A^2}]\\
		&\leq\log K_\alpha+\MI[\delta_{A^2}],
	\end{aligned}
	\label{eqn:MA_LB_2}
\end{equation}
where $\mathcal{H}(\cdot)$ denotes the entropy. The inequality follows from the monotonicity of the constant-modulus mutual information in transmit power, together with $\s\leq A^2$ almost surely. Combining \eqref{eqn:MA_LB_1} and \eqref{eqn:MA_LB_2}, and recalling that $\Kl\geq2$ yield \eqref{eqn:MA_Kl}.

Furthermore, as $\beta$ is increasing with $\alpha$, $\Kl$ is non-decreasing with $\alpha$.

\subsection{Upper Bound}

We first account for repeated sensing-information gains by combining the terms in $\phi(s)$ having identical positive $\sigma_\ell$. Let $\{\tilde{\sigma}_j\}_{j=1}^{\Lhseff}$ denote the distinct positive values among $\{\sigma_\ell:\lambda_\ell\sigma_\ell>0\}$, where $\Lhseff\in\{1,2\cdots,\Lhs\}$ is the number of these distinct values,  and define\footnote{A numerical example is that $\{\sigma_1,\sigma_2,\sigma_3,\sigma_4\}=\{4,2,2,0\}$ and $\{\lambda_1,\lambda_2,\lambda_3,\lambda_4\}=\{1,2,3,4\}$. Then, we have $\Lhseff=2$, $\{\tilde{\sigma}_1,\tilde{\sigma}_2\}=\{4,2\}$, and $\{\tilde{\lambda}_0,\tilde{\lambda}_1,\tilde{\lambda}_2\}=\{4,1,5\}$.}
\begin{equation*}
	\tilde{\lambda}_j :=
	\sum_{\ell:\sigma_\ell=\tilde{\sigma}_j}\lambda_\ell>0, \qquad \tilde{\lambda}_0=\sum_{\ell:\sigma_\ell=0}\lambda_\ell.
\end{equation*}
Thus, 
\begin{equation*}
	\phi(s)= \sum_{\ell=1}^{\Lhs}\frac{\lambda_\ell}{1+\sigma_\ell s} = \tilde{\lambda}_0+
	\sum_{j=1}^{\Lhseff}\frac{\tilde{\lambda}_j}{1+\tilde{\sigma}_js}.
\end{equation*}

By Proposition \ref{proposition:MA_KKT}, $l_\alpha(s;\Psopt)=0$ must hold at every optimal mass point, i.e.,
\begin{equation*}
	K_\alpha \leq \card\left(\left\{s:l_\alpha(s;\Psopt)=0,s\in[0,A^2]\right\}\right).
\end{equation*}
Since every interior mass point is a local maximizer of $l_\alpha(s;\Psopt)$, it is necessarily a zero of its derivative. Hence, the number of interior mass points is bounded by the number of zeros of $V_\alpha$, while the possible boundary mass point at $s=A^2$ contributes at most one additional point. By Rolle's theorem, we have
\begin{equation*}
	K_\alpha\leq
	1+\noz_{(0,A^2)}(l_\alpha'(s;\Psopt)),
	\label{eqn:MA_UB_Rolle}
\end{equation*}
where $\noz_\mathcal{A}(f)$ denotes the number of distinct zeros of function $f$ in interval $\mathcal{A}$. Define
\begin{equation*}
	Q(z)=\prod_{j=1}^{\Lhseff}(1+\tilde{\sigma}_jz)^2
\end{equation*}
and
\begin{equation}
	\begin{aligned}
		V_\alpha(z)
		=&\ Q(z)l'_\alpha(z;\Psopt)\\
		=&\ Q(z)\kappa'(z;\Psopt)-\beta\sum_{\ell=1}^{\Lhseff}
		\tilde{\lambda}_\ell\tilde{\sigma}_\ell
		\prod_{i\neq \ell}^{\Lhseff}(1+\tilde{\sigma}_i z)^2.
	\end{aligned}
	\label{eqn:Valpha_def}
\end{equation}
Since $\kappa(z;\Ps)$ admits an entire extension (as established in Appendix \ref{appendix:kappa_analyticity}), $V_\alpha(z)$ is entire. Moreover, $Q(s)>0$ for $s\geq0$, so $V_\alpha$ and $l_\alpha'$ have identical zeros on $(0,A^2)$. Hence, 
\begin{equation*}
	K_\alpha\leq
	1+\noz_{(0,A^2)}(V_\alpha).
\end{equation*}

For each $j\in\{1,\cdots,\Lhseff\}$, let
\begin{equation*}
	z_j:=-\frac{1}{\tilde{\sigma}_j},
	\qquad
	d_j:=\tilde{\lambda}_j\tilde{\sigma}_j
	\prod_{i\neq j}^{\Lhseff}
	\left(1-\frac{\tilde{\sigma}_i}{\tilde{\sigma}_j}\right)^2.
\end{equation*}
Evaluating \eqref{eqn:Valpha_def} at $z_j$, we obtain
\begin{equation*}
	V_\alpha(z_j)=-\beta d_j\neq0.
	\label{eqn:MA_H_nonzero}
\end{equation*}

\begin{figure}[t]
	\centering
	\includegraphics[scale=1]{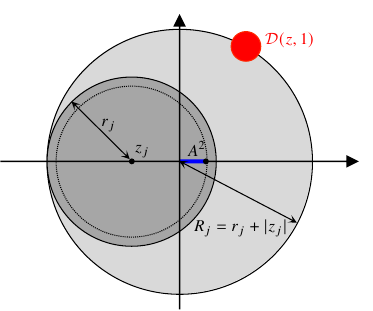}
	\caption{The Jensen's disks.}
	\label{fig:Jensen_disk}
\end{figure}

Consider the disk $\mathcal{D}(z_j,|z_j|+A^2)$, where $\mathcal{D}(c_0,r)$ is defined as the disk centered at $c_0$ with a radius $r$, i.e., $$\mathcal{D}(c_0,r):=\left\{z\in\mathbb{C}: |z-c_0| < r\right\}.$$
It's clear that the disk $\mathcal{D}(z_j,|z_j|+A^2)$ encloses the interval $(0,A^2)$, as shown in Fig.~\ref{fig:Jensen_disk}. Specifically, the region enclosed by the dashed circle is disk $\mathcal{D}(z_j,|z_j|+A^2)$, while the blue segment is the interval $(0,A^2)$. This implies
\begin{equation*}
	K_\alpha \leq 1+ \noz_{(0,A^2)}(V_\alpha) \leq 1+  \noz_{\mathcal{D}(z_j,|z_j|+A^2)}(V_\alpha).
\end{equation*}
We consider a larger disk $\mathcal{D}(z_j,r_j)$ with radius $r_j>|z_j|+A^2$. Applying the zero-counting estimate derived from Jensen's formula \cite[Eq. (D.1.2)]{dyatlov2019mathematical} gives\footnote{This estimate is a direct extension of the zero-counting bound in \cite{dyatlov2019mathematical}, which considers the special case where the outer radius is twice the inner radius. In fact, the argument applies to arbitrary concentric Jensen's discs with radii $|z_j|+A^2$ and $r_j>|z_j|+A^2$, yielding the denominator $\log(r_j/(|z_j|+A^2))$.}
\begin{equation}
	\begin{aligned}
		K_\alpha  
		&\leq 1+ \frac{
			\log\left(
			\frac{
				\max_{|z-z_j|\le r_j}|V_\alpha(z)|
			}{
				|V_\alpha(z_j)|
			}
			\right)
		}{
			\log\left(\frac{r_j}{|z_j|+A^2}\right)
		} \\
		&= 1+ \frac{\log(\max_{|z-z_j|\le r_j}|V_\alpha(z)|)-\log(-\beta d_j)}{\log(r_j)-\log(|z_j|+A^2)}.
	\end{aligned}
	\label{eqn:MA_Jensen}
\end{equation}

Next, we need to estimate $\max_{|z-z_j|\le r_j}|V_\alpha(z)|$. According to \eqref{eqn:Valpha_def}, we next find an upper bound for $|\kappa'(z;\Ps)|$.

Using the series representation \cite[Eq. (16.2.1)]{NIST_DLMF} of the hypergeometric function in \eqref{eq:radial_hypergeometric_pdf}, we obtain
\begin{equation*}
	p_{\mathsf{t}|\s}(t|z)
	=
	\mathrm{e}^{-\SNRc z}
	\sum_{k=0}^{\infty}
	\frac{(\SNRc z)^k}{k!}
	f_{\chi^2_{2(N+k)}(0)}(t),
\end{equation*}
where $	f_{\chi^2_{2(N+k)}(0)}(t)$ is the \ac{PDF} of a central Chi-squared distribution with degrees of freedom $2(N+k)$. 

Consider an even larger disk $\mathcal{D}(0,R_j)$ with $R_j:=r_j+|z_j|$, which completely encloses $\mathcal{D}(z_j,r_j)$, as shown in Fig. \ref{fig:Jensen_disk}. For $|z|<R_j$, the non-negativity of the series coefficients and the triangle inequality yield
\begin{equation*}
	\begin{aligned}
		|p_{\mathsf{t}|\s}(t|z)|
		&\le \mathrm{e}^{-\SNRc\Re\{z\}}
		\sum_{k=0}^{\infty}
		\frac{(\SNRc|z|)^k}{k!}	f_{\chi^2_{2(N+k)}(0)}(t)\\
		&=\mathrm{e}^{\SNRc(|z|-\Re\{z\})}
		p_{\mathsf{t}|\s}(t||z|)\\
		&\le \mathrm{e}^{2\SNRc R_j}p_{\mathsf{t}|\s}(t||z|).
	\end{aligned}
\end{equation*}

From \eqref{eq:gaussian_pdf_two_sided_bound} and by setting $M=A^2$, we have $
0\le-\log p_{\yc}(\yr;\Ps)
\le
N\log\pi+
\left(\|\yr\|+\sqrt{\SNRc}A\right)^2
=
N\log\pi+
(\sqrt{t/2}+\sqrt{\SNRc}A)^2.$ Therefore,
\begin{equation*}
	\begin{aligned}
		|\kappa(z;& P_s)|
		\le
		\mathrm{e}^{2\SNRc R_j} \!\!
		\int_0^\infty\!\! p_{\mathsf{t}|\s}(t||z|)
		\Big[
		N\log\pi+
		\Big(\sqrt{\frac t2}+\sqrt{\SNRc}A\Big)^2
		\Big]\diff t\\
		&= \mathrm{e}^{2\SNRc R_j}\Bigg[N\log\pi+\SNRc A^2\!+ \mathbb{E}\bigg[\frac{\mathsf{t}}{2}+\sqrt{2\SNRc}A\sqrt{\frac{\mathsf{t}}{2}} \bigg|\s=|z|\bigg]\Bigg]\!.
	\end{aligned}
\end{equation*}
Recall from \eqref{eqn:conditional_expectation_yc_square} that
\begin{equation*}
	\mathbb{E}\left[\frac{\mathsf{t}}{2}\bigg|\s=s\right] = \mathbb{E}\Big[\|\yc\|^2\mid\s=s\Big]  = N+\SNRc s,
\end{equation*}
and from \eqref{eqn:conditional_expectation_yc_Jensen} that
\begin{equation*}
	\mathbb{E}\left[\sqrt{\frac{\mathsf{t}}{2}}\bigg|\s=s\right]  =  \mathbb{E}\left[\|\yc\| \mid \s=s \right] \leq \sqrt{N+\SNRc s},
\end{equation*}
for $|z|<R_j$, we establish
\begin{equation*}
	\begin{aligned}
		|\kappa(z;P_s)| \leq \mathcal{B}(R_j) := \ \mathrm{e}^{2\SNRc R_j} 
		& \Big[N\log\pi+\SNRc A^2+ N+\SNRc R_j \\
		&\qquad + 2\sqrt{\SNRc}A\sqrt{N+\SNRc R_j}\Big].
	\end{aligned}
\end{equation*}
Consider the unit disk $\mathcal{D}(z,1)$, see the red disk in Fig.~\ref{fig:Jensen_disk}. Here, we plot the extreme case where $|z|=R_j$.

Cauchy's derivative estimate \cite[Ch.~2, Sec.~4, Corollary~4.3]{Stein2003Complex} gives
\begin{equation*}
	|\kappa'(z;\Ps)| \leq \max_{|z-w|\leq 1} |\kappa(w;\Ps)|\leq \mathcal{B}(R_j+1),
\end{equation*}
as $|w|\leq R_j+1$. Therefore, from \eqref{eqn:Valpha_def} and the triangular inequality,  we have
\begin{equation*}
	\begin{aligned}
		|V_\alpha(z)|  &\leq |Q(z)| |\kappa'(z;\Psopt)| + |\beta|\sum_{\ell=1}^{\Lhseff}
		\tilde{\lambda}_\ell\tilde{\sigma}_\ell
		\prod_{i\neq \ell}^{\Lhseff}|1+\tilde{\sigma}_i z|^2 \\
		&\leq Q(R_j)\mathcal{B}(R_j+1) + |\beta| \sum_{\ell=1}^{\Lhseff}
		\tilde{\lambda}_\ell\tilde{\sigma}_\ell
		\prod_{i\neq \ell}^{\Lhseff}(1+\tilde{\sigma}_i R_j)^2 \\
		&= a(R_j) + |\beta| b(R_j),
	\end{aligned}
\end{equation*}
where 
\begin{equation*}
	\begin{aligned}
		a(R) &:= Q(R)\mathcal{B}(R+1), \\
		b(R) &:= \sum_{\ell=1}^{\Lhseff}
		\tilde{\lambda}_\ell\tilde{\sigma}_\ell
		\prod_{i\neq \ell}^{\Lhseff}(1+\tilde{\sigma}_i R)^2.
	\end{aligned}
\end{equation*}
Finally, we have
\begin{equation*}
	\max_{|z-z_j|\leq r_j} |V_\alpha(z)| \leq a(R_j)+|\beta|b(R_j).
\end{equation*}
As this inequality holds for each $j\in \{1,\cdots,\Lhseff\}$ and $r_j>|z_j|+A^2$, according to the relationship in \eqref{eqn:MA_Jensen}, we can set the upper bound to be \eqref{eqn:MA_Ku}.

Furthermore, as $|\beta|$ is decreasing with $\alpha$, $\Ku$ is non-decreasing with $\alpha$.
\end{proof}
\section{Derivation of the Algorithm \ref{alg:MA_solver} under MA Constraint}
\label{appendix:MA_numerical_algorithm}

This appendix provides a complete derivation of the numerical algorithm in Algorithm~\ref{alg:MA_solver}. We derive the mutual-information functional, the probability and support-location updates, and the KKT-based optimality criterion.

\subsection{Computation of Mutual Information Based on Discrete Power Distribution}
\label{sec:compute_IPs}
In this subsection, we provide procedure to (numerically) compute $\MI[\Ps]$ when the input power distribution is discrete, e.g., when 
\begin{equation}
	\Ps=\sum_{k=1}^{K}p_k\delta_{s_k},
	\label{eqn:app_discrete_measure}
\end{equation}
where $0\leq s_k\leq A^2$, $p_k>0$ and $\sum_{k=1}^Kp_k=1$, is given. By variable transformation \eqref{eqn:t_transform}, for any feasible power distribution $\Ps$,
the unconditional scalar output PDF is
\begin{equation}
	p_{\mathsf{t}}(t;\Ps)
	=\int_0^{A^2} \!\! p_{\mathsf{t}|\s}(t|s)\,\diff \Ps(s) = \sum_{k=1}^{K}p_kp_{\mathsf{t}|\s}(t|s_k),
	\label{eq:app_discrete_mixture}
\end{equation}
where $p_{\mathsf{t}|\s}(t|s)$ can be calculated as \eqref{eq:app_scalar_pdf}. By definition,
\begin{equation}
	\begin{aligned}
		\MI[\Ps]
		&= \MI(\x;\yc) =h(\yc)-h(\zc)  \\
		&=\int_0^{A^2}\kappa(s;\Ps)\diff\Ps(s)
		-N\log(\pi \mathrm{e})\\
		&=\sum_{k=1}^{K}p_k\kappa(s_k;\Ps)
		-N\log(\pi \mathrm{e}),
	\end{aligned}
	\label{eqn:app_scalar_MI}
\end{equation}
where $\kappa(s;\Ps)$ can be numerically evaluated using its radial integral representation derived in Appendix \ref{appendix:MA_Ps}, given by
\begin{equation*}
	\begin{aligned}
		&\ \kappa(s;\Ps)
		 =N\log\pi+
		\\
		&\ \ \  \int_0^\infty p_{\mathsf{t}|\s}(t|s) \Bigg\{  \frac{t}{2}  -\log \Big[\sum_{k=1}^{K}p_k\mathrm{e}^{-\SNRc  s_k}
		{}_0F_1\left(;N;\frac{\SNRc  s_kt}{2}\right)\Big]\Bigg\}\diff t.
	\end{aligned}
\end{equation*}

\subsection{Finite-Dimensional Pareto Optimization}

Under the MA constraint, the optimization problem in \eqref{eqn:Pareto} is
\begin{equation}
	\begin{aligned}
		\max_{\Ps}\quad& \Fobj[\Ps]=
		\MI[\Ps]+\beta\xi[\Ps]\\
		\mathrm{s.t.}\quad&
		\supp(\Ps)\subseteq[0,A^2],
	\end{aligned}
	\label{eqn:app_MA_problem}
\end{equation}
where $\beta\leq0$ and $\xi[\Ps]
=\int_0^{A^2}\phi(s)\,\diff \Ps(s)$. In part \ref{sec:compute_IPs}, we provide the procedure to determine $\MI[\Ps]$ when $\Ps$ is discrete. We first introduce the ISAC Pareto utility density in the power domain
\begin{equation}
	f_\alpha(s;\Ps) = \kappa(s;\Ps)+\beta\phi(s)-N\log(\pi\mathrm{e}).
	\label{eqn:ISAC_pareto_utility_density}
\end{equation}
From \eqref{eqn:app_scalar_MI} and \eqref{eqn:app_MA_problem}, it is obvious that 
\begin{equation}
	\Fobj[\Ps]
	=\int_0^{A^2}
	f_\alpha(s;\Ps)\,\diff \Ps(s).
	\label{eqn:app_objective_density}
\end{equation}

By Theorem \ref{theorem:MA_discrete}, the optimal distribution possesses finitely many mass points. Therefore, consider the representation \eqref{eqn:app_discrete_measure}, $\Fobj[\Ps]$ may be rewritten as  $\Fobj[\boldsymbol{p},\boldsymbol{s}]$, where
\begin{equation*}
	\begin{aligned}
		\boldsymbol{p}&:=[p_1,\ldots,p_K]^\trans,\\
		\boldsymbol{s}&:=[s_1,\ldots,s_K]^\trans.
	\end{aligned}
\end{equation*}
For fixed $K$, the numerical optimization becomes
\begin{equation}
	\begin{aligned}
		\max_{\boldsymbol{p},\boldsymbol{s}}\quad&
		\Fobj[\boldsymbol{p},\boldsymbol{s}]\\
		\mathrm{s.t.}\quad&
		p_k\geq0,\quad k=1,\ldots,K,\\
		&\sum_{k=1}^{K}p_k=1,\\
		&0\leq s_k\leq A^2.
	\end{aligned}
	\label{eqn:app_finite_problem}
\end{equation}
Next, we will derive an alternating optimization procedure for $\boldsymbol{p}$ and $\boldsymbol{s}$.

\subsection{Alternating Optimization Procedure for Mass-Point Probabilities $\boldsymbol{p}$ and Mass-Point Locations $\boldsymbol{s}$}
\subsubsection{Optimization of Mass-Point Probabilities}

For fixed mass-point locations
$\boldsymbol{s}$, the optimization
in \eqref{eqn:app_finite_problem}
reduces to
\begin{equation}
	\begin{aligned}
		\max_{\boldsymbol{p}}\quad&
		\Fobj[\boldsymbol{p},\boldsymbol{s}]\\
		\mathrm{s.t.}\quad&
		\boldsymbol{p}\geq\mathbf0,\quad
		\mathbf1^\trans\boldsymbol{p}=1.
	\end{aligned}
	\label{eqn:app_probability_problem}
\end{equation}
Since the output entropy is concave in the output PDF, and $p_{\mathsf{t}}(t;\Ps)$ (see, e.g., in \eqref{eq:app_discrete_mixture}) is affine in $\boldsymbol{p}$, the objective is concave in $\boldsymbol{p}$. For a discrete distribution \eqref{eqn:app_discrete_measure}, \eqref{eqn:app_objective_density} becomes
\begin{equation*}
	\Fobj[\boldsymbol{p},\boldsymbol{s}]
	=\sum_{k=1}^{K}p_k
	f_\alpha(s_k;\Ps).
\end{equation*}

We now derive the BA update. Let the current probability vector be $\boldsymbol{p}^{[n]}$, with strictly positive entries, and denote its induced output
PDF by $p_{\mathsf{t}}^{[n]}(t) := p_{\mathsf{t}}(t;\boldsymbol{p}^{[n]})$, given as
\begin{equation*}
	p_{\mathsf{t}}^{[n]}(t)
	=\sum_{k=1}^{K}
	p_k^{[n]}p_{\mathsf{t}|\s }(t|s_k).
\end{equation*}
For any candidate probability vector $\boldsymbol{p}^{[n+1]}$ in the $(n+1)$-th iteration, its induced PDF is given by
\begin{equation*}
	p_{\mathsf{t}}^{[n+1]}(t)
	=\sum_{k=1}^{K}
	p_k^{[n+1]}p_{\mathsf{t}|\s }(t|s_k).
\end{equation*}
From \eqref{eqn:app_scalar_MI}:
\begin{equation*}
	\begin{aligned}
		\Fobj[& \boldsymbol{p}^{[n+1]},\boldsymbol{s}] \\
		&= \sum_{k=1}^Kp_k^{[n+1]}[\kappa(s_k;\Ps^{[n+1]})+\beta\phi(s_k)]-N\log(\pi\mathrm{e}) \\
		&= \sum_{k=1}^Kp_k^{[n+1]}[\kappa(s_k;\Ps^{[n]})+\beta\phi(s_k)-N\log(\pi\mathrm{e})]  \\
		&\ \ + \sum_{k=1}^K p_k^{[n+1]} [\kappa(s_k;\Ps^{[n+1]})-\kappa(s_k;\Ps^{[n]})] \\
		&=
		\sum_{k=1}^Kp_k^{[n+1]}
		f_\alpha(s_k;\Ps^{[n]})\\
		&\ \ -\sum_{k=1}^Kp_k^{[n+1]}
		\int_0^\infty
		p_{\mathsf t|\s}(t|s_k)
		\log\frac{p_{\mathsf t}^{[n+1]}(t)}
		{p_{\mathsf t}^{[n]}(t)}
		\,\diff t\\
		&=
		\sum_{k=1}^Kp_k^{[n+1]}
		f_\alpha(s_k;\Ps^{[n]})\\
		&\ \ -\int_0^\infty
		\underbrace{\left[
			\sum_{k=1}^Kp_k^{[n+1]}
			p_{\mathsf t|s}(t|s_k)
			\right]}_{=p_{\mathsf t}^{[n+1]}(t)}
		\log\frac{p_{\mathsf t}^{[n+1]}(t)}
		{p_{\mathsf t}^{[n]}(t)}
		\,\diff t\\
		&= \sum_{k=1}^Kp_k^{[n+1]}
		f_\alpha(s_k;\Ps^{[n]}) -
		D_{\mathrm{KL}}\left(
		p_{\mathsf t}^{[n+1]}
		\middle\|
		p_{\mathsf t}^{[n]}
		\right),
	\end{aligned}
\end{equation*}
where $D_{\mathrm{KL}}$ is the KL-divergence. From the data-processing inequality $	D_{\mathrm{KL}}(p_{\mathsf t}^{[n+1]}\|p_{\mathsf t}^{[n]})
\le
D_{\mathrm{KL}}(\boldsymbol{p}^{[n+1]}\|\boldsymbol{p}^{[n]})$, we have 
\begin{equation}
	\begin{aligned}
		\Fobj[\boldsymbol{p}^{[n+1]},\boldsymbol{s}]
		\ge{}&
		\sum_{k=1}^Kp_k^{[n+1]}
		f_\alpha(s_k;\Ps^{[n]})\\
		&\qquad -
		\sum_{k=1}^Kp_k^{[n+1]}
		\log\frac{p_k^{[n+1]}}{p_k^{[n]}},
	\end{aligned}
	\label{eq:app_BA_minorizer}
\end{equation}
The lower bound is tight at the current iterate, i.e., when $ \boldsymbol{p}^{[n+1]}=\boldsymbol{p}^{[n]} $. Maximizing this lower bound therefore yields a monotonically non-decreasing update of the Pareto objective. Thus, we aim to solve
\begin{equation*}
	\begin{aligned}
		&\max_{\boldsymbol{p}} &\quad &\sum_{k=1}^Kp_k
		f_\alpha(s_k;\Ps^{[n]})
		-\sum_{k=1}^Kp_k
		\log\frac{p_k}{p_k^{[n]}} \\
		&\mathrm{s.t.}& & p_k\geq 0, \quad 1\leq k\leq K,\\
		& & & \sum_{k=1}^K p_k = 1.
	\end{aligned}
\end{equation*}
Using the method of Lagrange multiplier, we thus obtain an update of $\boldsymbol{p}^{[n+1]}$ as
\begin{equation}
	p_k^{[n+1]}
	=
	\frac{
		p_k^{[n]}
		\exp\left\{f_\alpha(s_k;\Ps^{[n]})\right\}
	}{
		\displaystyle\sum_{j=1}^Kp_j^{[n]}
		\exp\left\{f_\alpha(s_j;\Ps^{[n]})\right\}
	}.
	\label{eq:MA_probability_update}
\end{equation}

Since the lower bound in \eqref{eq:app_BA_minorizer} is tight at the current iterate, this update satisfies
\begin{equation*}
	\Fobj[\boldsymbol{p}^{[n+1]},\boldsymbol{s}]
	\geq \Fobj[\boldsymbol{p}^{[n]},\boldsymbol{s}].
\end{equation*}
Hence, the probability update is monotonically non-decreasing in the objective.

\subsubsection{Optimization of Mass-Point Locations}

After updating the mass-point probabilities based on $\boldsymbol{s}^{[k]}$, we next optimize their locations to $\boldsymbol{s}^{[k+1]}$ while keeping the updated probabilities fixed. Specifically, given $\boldsymbol{p}^{[n+1]}$, we consider
\begin{equation}
	\begin{aligned}
		\max_{\boldsymbol{s}}\quad&
		\Fobj[\boldsymbol{p}^{[n+1]},\boldsymbol{s}]\\
		\mathrm{s.t.}\quad&
		0\le s_k\le A^2,\quad k=1,\ldots,K.
	\end{aligned}
	\label{eq:MA_mass_update}
\end{equation}
This problem can be (numerically) solved using common method such as gradient ascent. Let
\begin{equation*}
	\widetilde \Ps^{[n+1]}
	:=
	\sum_{k=1}^{K}
	p_k^{[n+1]}\delta_{s_k^{[n]}}
\end{equation*}
denote the intermediate distribution after the probability update but before the location update. After optimization, we will have
\begin{equation*}
	\Fobj[\Ps^{[n+1]}]
	\ge
	\Fobj[\widetilde \Ps^{[n+1]}]
	\ge
	\Fobj[\Ps^{[n]}].
\end{equation*}

Therefore, each alternating iteration is monotonically non-decreasing in the objective. The inner alternating optimization loop can be terminated when objective improvement falls below $\epsilon_{\mathrm{in}}$. Since the location optimization is generally non-concave, global optimality is subsequently assessed through the KKT-based optimality gap.

\subsection{KKT Conditions and Optimality Gap}

We next establish a numerical optimality criterion for the adaptive support refinement. Integrating the KKT conditions in Proposition \ref{proposition:MA_KKT} with respect to $\Psopt$ yields
\begin{equation}
	\begin{cases}
		f_\alpha(s;\Psopt)
		\le \Fobj[\Psopt],
		& s\in[0,A^2]\\[2mm]
		f_\alpha(s;\Psopt)
		=\Fobj[\Psopt],
		& s\in\supp(\Psopt).
	\end{cases}
	\label{eq:MA_numerical_KKT}
\end{equation}
These conditions are also sufficient for global optimality due to concavity. Motivated by \eqref{eq:MA_numerical_KKT}, for any feasible candidate distribution $\Ps$, define the optimality gap
\begin{equation}
	\Delta_\alpha[\Ps]
	:=
	\max_{s\in[0,A^2]}
	f_\alpha(s;\Ps)-\Fobj[\Ps].
	\label{eq:MA_KKT_gap}
\end{equation}
Since $\Fobj[\Ps]$ is the expectation of $f_\alpha(s;\Ps)$ under $\Ps$, it follows that $\Delta_\alpha[\Ps]\ge0$. Furthermore, for any feasible distribution $\mathbb{Q}_{\s}$, the concavity of $\Fobj$ yields
\begin{equation*}
	\begin{aligned}
		\Fobj[\mathbb{Q}_{\s}]-\Fobj[\Ps]
		&\le
		\int_0^{A^2}
		f_\alpha(s;\Ps)\,\diff \mathbb{Q}_{\s}(s)
		-\Fobj[\Ps]\\
		&\le
		\Delta_\alpha[\Ps].
	\end{aligned}
\end{equation*}
In particular, taking $\mathbb{Q}_{\s}=\Psopt$ gives 
\begin{equation}
	0\le
	\Fobj[\Psopt]-\Fobj[\Ps]
	\le\Delta_\alpha[\Ps].
	\label{eqn:KKT_inequality_MA}
\end{equation}
Hence, the KKT gap provides an upper bound on the global objective suboptimality. In particular, $\Delta_\alpha[\Ps]=0$ if and only if $\Ps$ is globally optimal.

\subsection{Adaptive Support Refinement}

Suppose that the current candidate distribution is $\Ps=\sum_{k=1}^{K}p_k\delta_{s_k}$. If $\Delta_\alpha[\Ps]>0$, then there exists at least one feasible power value $s_0$ such that
\begin{equation*}
	f_\alpha(s_0;\Ps) >\Fobj[\Ps].
\end{equation*}
Introducing an infinitesimal probability mass at $s_0$ strictly increases the objective to first order. The location yielding the largest first-order improvement
is therefore
\begin{equation}
	s_{\mathrm{new}}
	=\arg\max_{s\in[0,A^2]}
	f_\alpha(s;\Ps).
	\label{eq:app_support_insertion}
\end{equation}

To enlarge the support, consider the initialization
\begin{equation*}
	P_{\s,\mathrm{new}}
	=(1-\zeta)\Ps
	+ \zeta \delta_{s_{\mathrm{new}}},
\end{equation*}
where $0<\zeta<1$ is a small initialization probability.

The probability and location optimizations are then repeated over the enlarged support. In numerical implementations, mass points with sufficiently small probabilities may be removed, and nearly coincident mass points may be merged. Such modifications must be followed by reevaluation of the objective and KKT gap.

The adaptive procedure is terminated when
\begin{equation*}
	\Delta_\alpha[\Ps]
	\leq\epsilon_{\mathrm{out}}.
\end{equation*}

\section{Derivation of the Algorithm \ref{alg:GP_solver} under GP Constraint}
\label{appendix:GP_numerical_algorithm}

\subsection{Finite-Dimensional Approximation}
Consider the GP-constrained Pareto optimization in \eqref{eqn:Pareto}. Let $\{s_1,\ldots,s_K\}$ be a finite grid. The power distribution is approximated by
\[
\Ps=\sum_{k=1}^{K}p_k\delta_{s_k},
\]
where $\boldsymbol{p}=[p_1,\ldots,p_K]^\trans$ is a probability vector. Different than the MA case, $\boldsymbol{s}=[s_1,\cdots,s_K]$ is the preset mass point grid that is not an optimization variable. Practically, we need to set both $\max\{s_1,\cdots,s_K\}$ and $K$ sufficiently large to yield good approximation. The resulting optimization problem is
\[
\begin{aligned}
	\max_{\boldsymbol{p}}\quad&
	\Fobj[\boldsymbol{p}]
	=\MI[\boldsymbol{p}]+\beta\sum_{k=1}^{K}p_k\phi(s_k)\\
	\mathrm{s.t.}\quad&
	\sum_{k=1}^{K}p_k=1,\\
	&\sum_{k=1}^{K}p_ks_k\leq B,\\
	&p_k\geq0,\quad k=1,\cdots,K.
\end{aligned}
\]
It is clear that $\Fobj[\boldsymbol{p}]$ is a concave maximization over a convex feasible set. Consequently, any feasible solution satisfying its KKT conditions is globally optimal for the discretized problem.

\subsection{Probability Update}
Similar to derivations in Appendix \ref{appendix:MA_numerical_algorithm}, the probability update can be derived as
\begin{equation}
	p_k^{[n+1]}
	=
	\frac{
		p_k^{[n]}
		\exp\left\{f_\alpha(s_k;\Ps^{[n]})-\nu_\alpha s_k\right\}
	}{
		\displaystyle\sum_{j=1}^Kp_j^{[n]}
		\exp\left\{f_\alpha(s_j;\Ps^{[n]})-\nu_\alpha s_j\right\}
	},
	\label{eqn:GP_probability_update}
\end{equation}
where $f_\alpha(s;\Ps)$ follows the definition in \eqref{eqn:ISAC_pareto_utility_density}, with $\beta(\alpha)$ evaluated under the GP constraint according to \eqref{eqn:rate_opt_GP} and \eqref{eqn:E_opt_GP}. In \eqref{eqn:GP_probability_update}, $\nu_\alpha$ can be numerically obtained through one-dimensional search, and satisfies the slackness condition \eqref{eqn:KKT_GP_slack}:
\begin{equation}
	\nu_\alpha \left(\sum_{k=1}^{K} p_k s_k-N\rho\right)=0.
\end{equation}
Specifically, we first evaluate the probability update with $\nu_\alpha=0$. If the resulting probability distribution satisfies the average-power constraint, we retain $\nu_\alpha=0$. Otherwise, we determine $\nu_\alpha>0$ such that $\sum_{k=1}^{K}p_k^{[n+1]}s_k=N\rho$. This ensures that each probability update satisfies the average-power constraint and the complementary-slackness condition.

\subsection{KKT Conditions and Optimality Gap}

Following the same argument as in Appendix \ref{appendix:MA_numerical_algorithm}, the KKT-based optimality gap under GP is given by
\begin{equation}
	\Delta_\alpha[\Ps]
=
\max_{1\leq k\leq K}
\left\{
f_\alpha(s_k;\Ps)-\nu_\alpha s_k
\right\}
-\Fobj[\Ps]
+\nu_\alpha B.
\label{eqn:GP_KKT_gap}
\end{equation}
Analogously to \eqref{eqn:KKT_inequality_MA}, this gap bounds the objective suboptimality of the discretized problem. For numerical implementation, the supremum in \eqref{eqn:GP_KKT_gap} is approximated by the maximum over the prescribed power grid $\{s_k\}_{k=1}^{K}$, which is chosen to be sufficiently dense and to cover a sufficiently large power interval.

The procedure is terminated when
\begin{equation*}
	\Delta_\alpha[\Ps]
	\leq\epsilon.
\end{equation*}

\section{Proof of Proposition \ref{proposition:AOA_est}}
\label{appendix:AOA_est}
\begin{proof}
The Jacobian matrix is calculated as
\begin{equation*}
	\nabla_{\boldeta}\hs = \big[\mathsf{h}\dot{\v}(\uptheta), \v(\uptheta), \jmath \v(\uptheta)\big],
\end{equation*}
where $\dot{\v}:=\frac{\partial}{\partial \uptheta} \v(\uptheta)=\v(\uptheta)\odot\left([0,1,\cdots,\Lhs-1]^\trans \jmath\pi\cos(\uptheta)\right)$, with $\odot$ denoting the Hadamard product. It follows that
\begin{equation*}
	(\nabla_{\boldeta}\hs)^\herm(\nabla_{\boldeta}\hs)=\begin{bmatrix}
		|\mathsf{h}|^2 \dot{\v}^\herm\dot{\v} & \mathsf{h}^* \dot{\v}^\herm \v & \jmath\mathsf{h}^* \dot{\v}^\herm\v \\
		\mathsf{h}\v^\herm\dot{\v} & \v^\herm \v & \jmath \v^\herm\v \\
		-\jmath \mathsf{h} \v^\herm\dot{\v} & -\jmath \v^\herm \v & \v^\herm\v 
	\end{bmatrix}.
\end{equation*}
Further taking the real part and taking the expectation with respect to $\boldeta$ yield (here $\mathsf{h}$ and $\uptheta$ are independent, and $\mathsf{h}$ is zero-mean)
\begin{equation*}
		\mathbb{E}\left[\Re\{(\nabla_{\boldeta}\hs)^\herm(\nabla_{\boldeta}\hs)\}\right]
		=\diag\{a,1,1\},
\end{equation*}
where $\displaystyle a:=\sigma_{\mathsf h}^2 \pi^2\frac{(\Lhs-1)(2\Lhs-1)}{12}\big(1+\mathrm{e}^{-2\sigma_{\uptheta}^2}\cos(2\theta_0)\big)$ for simplicity. We have used the fact that 
$$\begin{aligned}
	\|\dot{\v}\|^2&=\frac{\sum_{\ell=0}^{\Lhs-1}\ell^2}{\Lhs}\pi^2\cos^2(\uptheta)\\
	&=\frac{(\Lhs-1)(2\Lhs-1)}6\pi^2\cos^2(\uptheta),
\end{aligned}$$
and $$\begin{aligned}
	\mathbb{E}_{\uptheta\sim\mathcal{N}(\theta_0,\sigma_{\uptheta}^2)}[\cos^2(\uptheta)] 
	=&\ \mathbb{E}_{\uptheta\sim\mathcal{N}(\theta_0,\sigma_{\uptheta}^2)}\left[\frac{1+\cos(2\uptheta)}2\right] \\
	=&\ \frac{1+\mathrm{e}^{-2\sigma_{\uptheta}^2}\cos(2\theta_0)}2,
\end{aligned} $$
where, in the last step, the characteristic function of Gaussian random variables is leveraged. Next, we can calculate
$$\Fp=\diag\left\{\frac1{\sigma_{\uptheta}^2},\frac2{\sigma_{\mathsf h}^2},\frac2{\sigma_{\mathsf h}^2}\right\}.$$
Based on simple matrix algebras, we may find that
\begin{equation*}
	\boldsymbol{G} = \diag\left\{\frac{2\sigma_{\uptheta}^2}{\noisevar}a,\frac{\sigma_{\mathsf h}^2}{\noisevar},\frac{\sigma_{\mathsf h}^2}{\noisevar}\right\} = \I_3 \diag\{\sigma_1,\sigma_2,\sigma_3\}\I_3^\trans, 
\end{equation*}
where we have chosen $\boldsymbol{Q} = \I_3$, $\sigma_1 = \frac{2\sigma_{\uptheta}^2}{\noisevar}a$, and $\sigma_2=\sigma_3=\frac{\sigma_{\mathsf h}^2}{\noisevar}$. Furthermore, $\boldLambda = \diag\{\sigma_{\uptheta}^2,0,0\}$ when $\W=\diag\{1,0,0\}$ is chosen. It follows immediately that $\lambda_1=\sigma_{\uptheta}^2$, $\lambda_2=\lambda_3=0$. 
\end{proof}

\balance
\bibliographystyle{IEEEtran}
\bibliography{ref}
\begin{IEEEbiographynophoto}{Zi-Jie Wang}
	(Member, IEEE) received the Bachelor's degree in electronic and computer engineering from Shanghai Jiao Tong University, Shanghai, China, in 2020, the Master's degree in electronic engineering (signal processing) from Nanyang Technological University, Singapore, in 2022, and the joint Ph.D. degree in information and communications engineering from Shanghai Jiao Tong University and the National University of Singapore, Singapore, in 2026. 
	Dr. Wang's research spans a broad range in signal processing and information theory, with their applications in wireless sensing, communications, and integrated sensing and communications (ISAC) systems. His research works have been published in \textsc{IEEE Transactions on Information Theory}, \textsc{IEEE Journal on Selected Areas in Communications}, etc.
\end{IEEEbiographynophoto}

\end{document}